\documentclass[journal=jacsat,manuscript=article]{achemso}

\usepackage[version=3]{mhchem} 
\usepackage{booktabs}
\usepackage{comment}
\usepackage{xcolor}
\usepackage{tcolorbox}
\SectionNumbersOn
\newcommand\RV[1]{\textcolor{black}{#1}}
\newcommand\RVt[1]{\textcolor{black}{#1}}
\newcommand\Lex[1]{\textcolor{black}{#1}}
\newcommand\second[1]{\textcolor{black}{#1}}

\author{Qinghao Shen}
\email{shen@differ.nl}
\affiliation[Unknown University]
{Dutch Institute for Fundamental Energy Research, Eindhoven, The Netherlands}
\alsoaffiliation {Department of Applied Physics, Eindhoven Institute of Renewable Energy Systems, Eindhoven University of Technology, Eindhoven, The Netherlands}

\author{Cas van Deursen} 
\affiliation[Unknown University]
{Dutch Institute for Fundamental Energy Research, Eindhoven, The Netherlands}
\alsoaffiliation {Department of Applied Physics, Eindhoven Institute of Renewable Energy Systems, Eindhoven University of Technology, Eindhoven, The Netherlands}

\author{Pieter Willem Groen} 
\affiliation[Unknown University]
{Dutch Institute for Fundamental Energy Research, Eindhoven, The Netherlands}

\author{Lex Kuijpers} 
\affiliation[Unknown University]
{Dutch Institute for Fundamental Energy Research, Eindhoven, The Netherlands}
\alsoaffiliation {Department of Applied Physics, Eindhoven Institute of Renewable Energy Systems, Eindhoven University of Technology, Eindhoven, The Netherlands}

\author{Mauritius C.M. van de Sanden}
\email{m.c.m.v.d.sanden@tue.nl}
\affiliation[Unknown University]
{Dutch Institute for Fundamental Energy Research, Eindhoven, The Netherlands}
\alsoaffiliation {Department of Applied Physics, Eindhoven Institute of Renewable Energy Systems, Eindhoven University of Technology, Eindhoven, The Netherlands}

\title[An \textsf{achemso} demo]
  {Flow–thermochemistry coupling governs pressure-dependent CO$_2$ conversion \RVt{in vortex-stabilized microwave plasma reactors}: Insights from three–dimensional CFD modeling}

\abbreviations{IR,NMR,UV}

\begin{document}

\begin{abstract}

\RVt{In this work, a three-dimensional computational fluid dynamics model is developed for a vortex-stabilized microwave CO$_2$ plasma reactor operating over the pressure range of 100--400~mbar. The model combines experimentally constrained, emission-based plasma sizes and volumetric heat-source distributions with thermally dominated finite-rate heavy-particle chemistry for a multi-component mixture. Turbulent flow and transport are described using the SST \(k\)-\(\omega\) model.}

\RVt{The model reproduces the measured radial gas-temperature profiles in the plasma core and the non-monotonic pressure dependence of CO$_2$ conversion, including a maximum at 150~mbar and a pronounced decrease at 400~mbar. A vortex-driven recirculation region redistributes gas upstream. Turbulent mixing and cooling are strongest near the upper reactor boundary, but their contribution decreases as pressure increases. The pressure dependence of conversion is determined by the competition between CO$_2$ dissociation and CO recombination. CO$_2$ direct dissociation reaction dominates in the high-temperature plasma core, whereas O-assisted conversion reaction contributes near the plasma edges and in the surrounding hot region. At 150~mbar, enhanced CO$_2$ dissociation is accompanied by limited CO loss, resulting in the highest conversion. With pressure increasing to 400~mbar, slower cooling and more frequent three-body collisions promote CO recombination in the afterglow, causing more than 60\% of the CO formed near the plasma to be lost downstream. Moreover, additional CO loss occurs in the upper region of the reactor at higher pressures because of the reduced cooling rate.}

\end{abstract}




\section{Introduction}

Carbon dioxide (CO$_2$), the primary anthropogenic greenhouse gas, has a long atmospheric lifetime and has become the dominant driver of global climate change as a result of extensive fossil fuel use since the 19th century. Despite the rapid growth of renewable energy in recent years, fossil fuels still account for more than 80\% of the global energy supply, creating a substantial gap between current energy trajectories and net-zero emission pathways \cite{global_energy2025}. This situation underscores the urgent need for effective technologies that can both valorize surplus renewable electricity and enable deep decarbonization of industrial processes \cite{di2018decarbonization, sun2024plasma}. Among emerging solutions, plasma-assisted CO$_2$ conversion has attracted increasing attention due to its unique advantages, including direct gas-phase activation under mild operating conditions, fast response to intermittent renewable power, and modular scalability, making it a promising pathway for sustainable energy and chemical production \cite{bogaerts2025plasma, hecimovic2024benchmarking,ong2022application}.

The \Lex{thermal} dissociation of CO$_2$ is highly endothermic, requires significantly elevated temperatures to proceed, and can be described by the following reaction:
\begin{equation}
    \textbf{CO$_2$}\rightarrow \textbf{CO} + \frac{1}{2}\textbf{O$_2$}, \hspace{1.2cm} \Delta H=2.93 \text{ eV}
\end{equation}
where the standard reaction enthalpy $\Delta H = 2.93$~eV represents the minimum net energy required to produce one CO molecule from CO$_2$ in a stable product mixture. Various plasma types have been investigated for CO$_2$ splitting, including dielectric barrier discharges (DBD), microwave (MW) discharges, gliding arcs (GA), nanosecond-pulsed discharges, as well as corona and spark discharges \cite{bongers2017plasma, snoeckx2017plasma, liu2026synchronous, guerra2022plasmas}. Among these, warm plasmas, such as GAs and MW discharges operated under specific conditions, generally exhibit significantly higher performance than cold plasmas (\textit{e.g.}, DBDs) \cite{bogaerts2018plasma,zhu2025situ}. Notably, the highest reported energy efficiency for CO$_2$ splitting—up to 90\% at a conversion of 10\%—was achieved in the 1980s using MW plasmas operated under supersonic gas flow conditions at pressures of 100--200~Torr \cite{fridman2008plasma}. This exceptionally high energy efficiency has often been attributed to vibrational enhancement. In MW plasmas, the reduced electric field typically lies in the range of 10--100~Td \cite{viegas2021resolving, viegas2020insight}, where a substantial fraction of the electron energy can be preferentially channeled into vibrational excitation \cite{pietanza2021advances}. However, such performance has never been reproducibly achieved. \Lex{Van de Steeg} \textit{et al.} reported that vibrational non-equilibrium in CO$_2$ MW plasmas remains very limited even at pressures as low as 60~mbar \cite{van2021redefining}. \second{Furthermore, Vialetto \textit{et al.} also reported that above 100~mbar, CO$_2$ dissociation is primarily driven by thermal reactions due to the high gas temperatures  \cite{vialetto2022charged}, rather than by electron impact, which plays a more crucial role at 60 mbar \cite{viegas2020insight}.}

\RVt{Recent experimental research has focused on improving MW plasma-based CO$_2$ conversion through reactor design, chemical promotion, and thermal management \cite{yang2025co2}. } \RVt{One strategy is to strengthen post-plasma cooling by reactor and flow-field optimization. Van Deursen \textit{et al.} showed that a converging--diverging nozzle improved conversion from about 10\% to 17\% at 900 mbar\cite{van2024effluent}. Hecimovic \textit{et al.} demonstrated that four cooled effluent channels can give CO$_2$ conversions up to 57\% at 900~mbar, while the best energy efficiency reached about 30\% at a lower conversion of 12\% \cite{hecimovic2023fast}.}

\RVt{A second strategy is chemical promotion. Kuijpers \textit{et al.} reported that CH$_4$ addition increases CO$_2$ conversion, CO yield, and fuel efficiency; the best case with 30\% CH$_4$ reached a total conversion of 50\% and a fuel efficiency above 70\% \cite{kuijpers2025microwave}. Biondo \textit{et al.} showed that coupling the plasma to a post-plasma carbon bed increased CO$_2$ conversion to above 40\% and reduced the energy cost below 2.8~eV~molecule$^{-1}$, corresponding to more than a fourfold increase in conversion and an almost fourfold decrease in energy cost compared with plasma operation without a carbon bed \cite{biondo2025coupling}.}

\RVt{A third strategy is thermal management through inlet-gas preheating. Mercer \textit{et al.} externally preheated the inlet CO$_2$ to evaluate the potential benefit of heat recycling in a 915~MHz MW plasma reactor \cite{mercer2025preheating}. Their results showed that, at 700~mbar and 1132~W, recycling an equivalent amount of heat corresponding to less than 10\% of the input power would be sufficient to increase the energy efficiency by up to a factor of 1.7.}

\RVt{However, a remaining challenge is to translate promising laboratory-scale performance into reactor concepts that are relevant for scale-up and process integration. Yang and Murphy recently emphasized that the technology readiness level of plasma CO$_2$ conversion remains low, and that progress towards industrialisation requires improved reactor design, consistent performance metrics, techno-economic assessment, and integration of plasma reactors into broader chemical processes \cite{yang2025co2}. Reactor-scale modeling can contribute to this development by linking reactor geometry, gas injection, recirculation, heat transfer, species transport, and finite-rate chemistry to measurable conversion and energy efficiency.} Viegas \textit{et al.} introduced a zero-dimensional Monte Carlo flux model focusing on the pressure-dependent contraction dynamics of MW plasmas \cite{viegas2020insight}. Subsequently, Vialetto \textit{et al.} investigated the kinetics and transport phenomena in CO$_2$ MW plasmas using a one-dimensional radial fluid model, which was validated against spatially resolved experimental measurements \cite{vialetto2022charged}. Kotov \textit{et al.} further developed a 1.5D laminar model with particular emphasis on thermochemical kinetics \cite{kotov2023validation}. However, a major limitation of these reduced-dimensional models is their inability to capture accurate gas flow phenomena, \Lex{such as recirculation}. In \Lex{vortex stabilized} MW plasma systems, gas recirculation within the plasma core plays a crucial role in determining the plasma shape, temperature distribution, species transport, local residence times, and mixing, thereby strongly influencing plasma stability and overall performance  \cite{van2025influence}. Consequently, a comprehensive CFD model \Lex{including the chemical kinetics} is required to obtain a more accurate and physically complete understanding of CO$_2$ dissociation in MW plasma reactors.

To the best of our knowledge, only a few studies have addressed CFD modeling of plasma-assisted gas conversion, primarily due to its inherent complexity. \RVt{Most existing studies simplify the problem either by reducing the dimensionality, often through two-dimensional axisymmetric descriptions, or by neglecting chemical kinetics and/or turbulence. Such reduced-dimensional models have provided important insight, but they cannot directly resolve non-axisymmetric features such as discrete tangential gas injection, three-dimensional vortex development, and azimuthal variations in recirculation. Ruijzendaal \textit{et al.} recently inferred upstream recirculation, downstream flow restriction, and bypass flow in a vortex-stabilized CO$_2$ microwave plasma, showing that global flow parameters alone are insufficient to describe the reactor transport \cite{ruijzendaal2026flow}. This motivates three-dimensional reactor-scale modeling with coupled flow, heat transfer, and chemistry.} 
\RVt{Mercer \textit{et al.} employed a three-dimensional CFD model to demonstrate the influence of a convergent--divergent nozzle on the flow field and temperature distribution at 700~mbar during post-plasma afterglow \cite{mercer2023post}.} Groen \textit{et al.} developed a 3D CFD model for a forward-vortex-flow MW CO$_2$ plasma system operating at intermediate pressures (60--250~mbar) \cite{groen2025modelling}. \RVt{To keep the simulations computationally tractable, both studies represented the plasma mixture as a single effective species. Such models provide useful insight into the hydrodynamic structure of vortex-stabilized MW plasma reactors, but the use of effective transport properties can oversimplify transport in highly non-isothermal reactive plasmas. This limitation is especially relevant because local-chemical-equilibrium-based transport properties may overestimate thermal conductivity and thereby affect the predicted size of the plasma-heated region.}

\RVt{In contrast to single-species CFD descriptions, the present 3D model solves finite-rate chemistry for a multi-component mixture. The local thermophysical and transport properties are evaluated from the kinetically determined mixture composition. Turbulence is described using the SST \(k\)-\(\omega\) model, allowing the influence of turbulent momentum, heat, and species transport to be included in the reactor-scale simulation. This treatment directly couples chemical conversion, species redistribution, heat transport, turbulence, and three-dimensional vortex-flow dynamics. The plasma is treated as a heat source derived from measured plasma-emission profiles, which reduces the computational cost while retaining the measured pressure-dependent plasma sizes. Radial gas-temperature profiles and post-plasma conversion data are used for validation. The validated model is then applied to determine how pressure changes the balance between CO$_2$ dissociation, CO recombination, flow recirculation, and turbulence behavior.}

\section{Experimental setup}

\begin{figure}[h]
\centering
\includegraphics[width=0.5\linewidth]{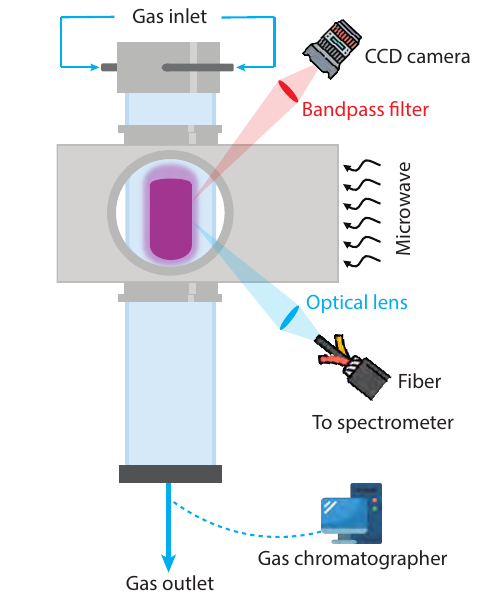}
\caption{Diagram of the experimental setup and the diagnostics used in this work.}
\label{fig:Diagram}
\end{figure}

A schematic of the experimental setup is shown in Fig.~\ref{fig:Diagram}. A MW discharge operated in a forward-vortex (FV) flow configuration was employed, following a reactor design concept previously developed for CO$_2$ dissociation studies \cite{van2024effluent}.  \second{The magnetron generates an adjustable continuous-wave, which is transmitted through a WR340 rectangular waveguide operating in the transverse electric TE$_{10}$ mode. A standing wave is established, such that the electric field is aligned parallel to the overall gas flow direction. A sliding short circuit is positioned at a quarter wavelength from the center of the quartz tube to maximize the electric field at the plasma location.} A circulator was installed to protect the MW source from reflected power, while an automatic three-stub tuner (HOMER S-TEAM STHT2450) was used to minimize power reflection. The applied MW power was maintained at 1000~W for all experiments.

The plasma was sustained inside a quartz tube (inner diameter: 27~mm; outer diameter: 30~mm) positioned at the center of the waveguide.  The feed gas was introduced through two tangential inlets \Lex{with 1 mm diameter} at the top of the reactor, generating a stabilizing FV flow. This hydrodynamic configuration confines the plasma column along the central axis, effectively preventing plasma--wall contact and minimizing degradation of the quartz tube \cite{mercer2025preheating}. Pure CO$_2$ was used as the feed gas, with a fixed total flow rate of 7 standard liters per minute (slm). The experiments were conducted at absolute pressures ranging from 100 to 400 mbar, regulated by a vacuum pump in combination with a back-pressure controller. Optical access for plasma diagnostics was provided by two cylindrical microwave-cutoff tubes mounted laterally on the waveguide.

The spatially resolved gas temperature in the plasma core was determined from the Doppler broadening of the O~777~nm triplet, corresponding to the \(^5\mathrm{P} \rightarrow {}^5\mathrm{S}\) transition. Each of the three individual lines in the triplet was independently fitted with a Voigt profile to extract the Gaussian (Doppler) component of the linewidth. The gas temperature was subsequently derived from the average Doppler \second{broadening} obtained from the three lines \cite{wolf2019characterization}.

\RVt{The spatial envelope of the plasma was obtained from O~777~nm emission images recorded with a CCD camera. The line-of-sight emission intensity was Abel-inverted to reconstruct the axisymmetric radial emission distribution. The plasma radius and length were then extracted from the resulting emission envelope. This emission-based definition follows the approach used in our previous studies of pressure-dependent CO$_2$ microwave plasma contraction \cite{wolf2019characterization,wolf2020implications}.}

\RVt{Previous studies have shown that electron-density profiles can be broader than O~777~nm emission envelopes \cite{viegas2021resolving}. No fixed broadening factor is applied here, because the relative widths of emission, electron density, and heat deposition depend on pressure, discharge mode, and local plasma chemistry. The agreement between the simulated gas-temperature profiles, CO$_2$ conversion, and experimental measurements suggests that this mapping is adequate for the present reactor-scale thermochemical CFD model, although uncertainty remains in the detailed power-deposition width. Future multi-wavelength imaging or self-consistent electromagnetic--CFD coupling would be needed to refine the heat-source distribution.}

The composition of the post-plasma effluent was analyzed using gas chromatography (GC, CompactGC 4.0). The performance of the \ce{CO2} dissociation process was evaluated primarily using two metrics: \ce{CO2} conversion ($\alpha$) [\%] and energy efficiency ($\eta$) [\%], \Lex{defined as follows}:
\begin{equation}
   \alpha= \frac{X_{CO}}{X_{CO}+X_{CO_2}}\times100\%
\end{equation}
\begin{equation}
   \eta= \alpha \frac{\Delta H_f}{ \mathrm{SEI}}
\end{equation}
where \Lex{\(X_{CO}\) and \(X_{CO_2}\) (both \RVt{dimensionless}) are the molar fractions of CO and CO$_2$}, \(\Delta H_f\) is the formation enthalpy of CO, which equals 2.93 eV, SEI [eV] is the specific energy input, which is defined as:
\begin{equation}
  \mathrm{SEI}=   \frac{P_{abs}}{ef_{in}}
\end{equation}
where \(P_{abs}\) [W] is the observed input power, $e$ is the elementary charge, \(f_{in}\) [mole s$^{-1}$]  is \RVt{the inflow rate}.

\section{Model description}

\begin{figure}[ht]
\centering
\includegraphics[width=1\linewidth]{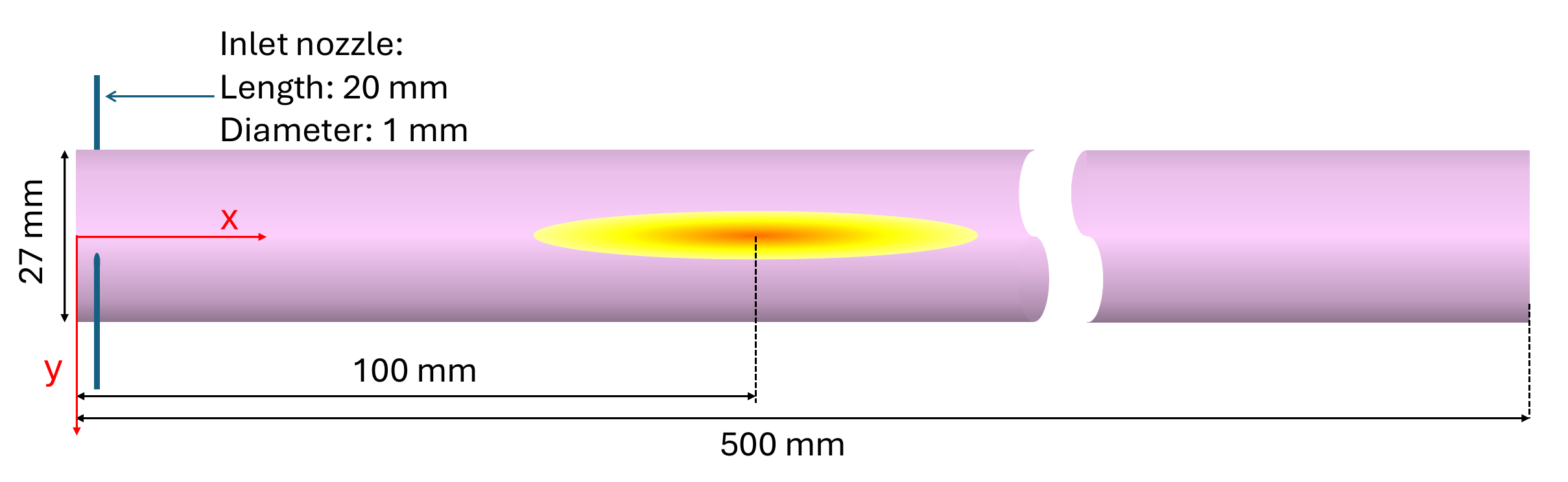}
\caption{The size of different parts in the reactor as implemented in the CFD model. \RV{The physical discharge tube is mounted vertically in the experiments, and the horizontal orientation is used only for visualization.}}
\label{fig:model_size}
\end{figure}

\RVt{A schematic of the CFD geometry is shown in Fig.~\ref{fig:model_size}.} \RV{The use of a prescribed heat source is supported by Raman-scattering measurements performed in the same forward-vortex MW CO$_2$ plasma reactor by Van de Steeg \textit{et al.}, which showed that vibrational non-equilibrium is already limited at 60~mbar \cite{van2021redefining}. Since the present simulations cover the higher pressure range of 100--400~mbar, where collision frequencies and vibrational-translational relaxation rates are higher, vibrational non-equilibrium is expected to be even less important for the overall CO$_2$ conversion. In addition, previous plasma-kinetic modeling showed that direct electron-impact CO$_2$ dissociation is at least two orders of magnitude weaker than thermal dissociation at 100~mbar and becomes relevant mainly at lower pressures around 60~mbar \cite{vialetto2022charged, viegas2020insight}. Based on these results, the present model treats the plasma as an experimentally constrained volumetric heat source and resolves the subsequent heavy-particle thermochemistry, species transport, heat transfer, turbulence, and three-dimensional flow.} \RV{This treatment still does not resolve electron-impact reactions, state-specific vibrational populations, charged-particle transport, or self-consistent microwave power deposition. These effects should be addressed in future coupled plasma-CFD models.}

\Lex{The three-dimensional simulations were performed using ANSYS Fluent \cite{matsson2022introduction}. \RVt{Because this work focuses on the final quasi-steady operating state, the time-derivative terms in the governing equations are zero. This steady-state formulation effectively reduces the computational cost while retaining the accuracy needed for the final reactor performance considered here.} In the selection of turbulence models, the SST (Shear Stress Transport) $k\!-\!\omega$ model is widely adopted due to its unique hybrid characteristics  \cite{mercer2023post}. \RVt{The SST model follows Menter's formulation, in which the $k$--$\omega$ model is retained in the inner region of the boundary layer and blended with a $k$--$\varepsilon$ formulation away from the wall \cite{menter1994two}. In the near-wall region, the transport equations for the turbulent kinetic energy $k$ and the specific dissipation rate $\omega$ are solved in the wall-adjacent cells. When the near-wall mesh is sufficiently refined, this treatment resolves the viscous sublayer directly and avoids the use of empirical wall functions.} Due to its balanced accuracy, stability, and broad applicability, in this work, the SST $k\!-\!\omega$ model is used for turbulence closure.}  

{
\color{black}
\subsection{Key governing equations}
\subsubsection{Fluid dynamics}

The turbulent gas flow behavior is described using a Reynolds-Averaged Navier-Stokes (RANS) approach that solves the following mass continuity and momentum continuity equations \cite{batchelor2000introduction}:
\begin{equation}
    \nabla\cdot(\rho\,\mathbf{u}) = 0
\end{equation}
\begin{equation}
    \label{eq: momemtum}
    \nabla \cdot (\rho\,\mathbf{u}\otimes\mathbf{u})
    = -\nabla p
    + \nabla\cdot\left[
        \mu\left(
            \nabla\mathbf{u} + (\nabla\mathbf{u})^\mathrm{T}
            - \tfrac{2}{3}(\nabla\cdot\mathbf{u})\,\mathbf{I}
        \right) -(\rho\,\overline{\mathbf{u}'\otimes\mathbf{u}'})
    \right]
\end{equation}
where \(\mathbf{u}\) and  \(\mathbf{u}'\) (both in [m s$^{-1}$]) are the \Lex{Reynolds averaged} flow and velocity and fluctuating velocity components, \(\rho\) [kg m$^{-3}$] denotes the mass density, \(p\) [Pa] is the gas pressure, \(\mathbf{I}\) is the unity tensor. \(\mu\) [Pa\,s] is the dynamic viscosity, and
\(\overline{\mathbf{u}'\otimes\mathbf{u}'}\) is the Reynolds-stress tensor. Both quantities are further described in the Supporting Information.

\subsubsection{Transport of species}

The local mass fraction of each species \(Y_i\) is described by the conservation of mass equation:
\begin{equation}
\nabla \cdot (\rho \, \mathbf{u} \, Y_i) = - \nabla \cdot \mathbf{J}_i + M_{i} R_i^{net},
\label{eq:local_mass_fraction}
\end{equation}
where \(R_i^{net}\) [mol m$^{-3}$s$^{-1}$] and \(M_{i}\) [kg~mol$^{-1}$] are the net rate and molecular weight of production of species \textit{i} by chemical reactions, respectively;  \(\mathbf{J}_i\) [kg m$^{-2}$s$^{-1}$] is the diffusion flux of species \(i\), arising due to gradients of concentration and temperature, which is calculated by:
\begin{equation}
\mathbf{J}_i = - \rho\left(  D_{i,m} + D_t \right) \nabla Y_i
             - D_{T_i} \frac{\nabla T}{T},
\label{eq:diffusion_flux}
\end{equation}
\RVt{where \(Y_i\) [dimensionless] is the mass fraction of species \(i\), \(D_{i,m}\) [m$^2$s$^{-1}$] is the mixture-averaged molecular diffusivity, and \(D_t\) [m$^2$ s$^{-1}$] is the turbulent diffusivity. These diffusivities are calculated as}
\begin{equation}
D_{i,m} = \; 
\Biggl( \sum_{j,\, j\neq i} \frac{X_j}{D_{ij}} 
       + \frac{X_i}{1-Y_i} \sum_{j,\, j\neq i} \frac{Y_j}{D_{ij}} 
\Biggr)^{-1}
\label{laminar_diffusive}
\end{equation}

\begin{equation}
D_{t} = \frac{\mu_t }{\rho Sc_t}
\label{turbulent_diffusive}
\end{equation}
\RVt{where \(X_i\) [dimensionless] is the mole fraction of species \(i\), \(D_{ij}\) [m$^2$ s$^{-1}$] is the binary diffusion coefficient of species \(i\) in species \(j\), \(Sc_t\) [dimensionless] is the turbulent Schmidt number, set to 0.7, and \(\mu_t\) [Pa\,s] is the turbulent viscosity.  More detailed information on \(D_{ij}\) and \(\mu_t\) is provided in the Supporting Information.}

\RVt{The thermal diffusion term \(D_{T_i}\) \RVt{[kg m$^{-1}$s$^{-1}$]} in Eq.(\ref{eq:diffusion_flux}) is included to account for species transport driven by temperature gradients. This effect may be relevant in the present reactor because steep temperature and composition gradients coexist near the plasma periphery and in the surrounding colder gas. Its inclusion therefore provides a more complete description of multi-component species transport than concentration-gradient diffusion alone. In this work, however, the thermal-diffusion contribution is not separately decomposed from the total species flux. A dedicated flux-budget analysis separating convective transport, molecular diffusion, turbulent diffusion, and thermal diffusion will be carried out in future work.} More detailed information about \(D_{T_i}\) can be found in the Supporting Information.

\subsubsection{Heat balance equation}
The total energy of a fluid is described by: 
\begin{equation}
\nabla \cdot \Bigg[ \rho \, \mathbf{u} \Big( h + \frac{\mathbf{u}\cdot\mathbf{u}}{2} \Big) \Bigg]
=
\nabla \cdot \Big(
k_{eff} \nabla T
- \sum_i h_i \mathbf{J}_i
+ \bar{\tau} \cdot \mathbf{u}
\Big)
+ Q_{(x, y, z)}
\label{eq:energy_equation}
\end{equation}
where \(h_i\) and \(h\) [J kg$^{-1}$] are the specific enthalpy of species \(i\) and the mixture specific enthalpy, respectively; \(\bar{\tau}\) [Pa] is the viscous stress tensor. The terms in parentheses on the right-hand side of Eq.(\ref{eq:energy_equation}) denote energy transfer due to conduction, species diffusion, and viscous dissipation, respectively \cite{matsson2022introduction}. \(Q_{(x, y, z)}\) [W m$^{-3}$]  is the local volumetric heat source, \(k_{eff}\) [W~m$^{-1}$~K$^{-1}$] is the effective thermal conductivity, defined as the sum of the laminar and turbulent contributions, \(k_{eff}=k_m+k_t\). These two thermal conductivities are calculated by \cite{shen2026boosting}:
\begin{equation}
k_t = \frac{c_p \, \mu_t}{Pr_t},
\label{eq:turbulent_conductivity}
\end{equation}
\begin{equation}
  k_{m}=\sum_i \frac{X_ik_i}{\sum_j X_j \phi_{ij}} 
  \label{eq:laminar_conductivity}
\end{equation}
where \(\phi_{ij}\) [dimensionless] is a interaction parameter that accounts for the effect of species \(j\) on the contribution of species \(i\) to the mixture thermal conductivity, \(c_p\) [J kg$^{-1}$K$^{-1}$] is the heat capacity of gas mixture,  \(Pr_{t}\) [dimensionless] is the turbulent Prandtl number, fixed at 0.85 \cite{matsson2022introduction}. More detailed information about \textit{h} , \(\bar{\tau}\), \(k_i\), and \(Q_{(x, y, z)}\) can be found in the Supporting Information.

\subsection{Boundary conditions and mesh design}

The velocity field satisfies a no-slip boundary condition on all walls, prescribing zero velocity at the fluid–solid interfaces. Both inlets are specified with mass flow rate boundary conditions, allowing the inlet pressure and velocity fields to be automatically determined by Fluent under varying pressure conditions. The inlet gas temperature is fixed at 300 K. At the reactor walls, zero normal diffusive flux is imposed for all species. The outlet is treated as a pressure outlet \cite{groen2025modelling}, while zero-gradient boundary conditions are applied to the species mass fractions. For heat transfer, the outlet boundary is treated with a zero normal temperature-gradient condition, so that heat leaves the domain mainly by convective outflow without imposing a fixed downstream temperature. Energy loss by the \RVt{quartz} boundary via convection and radiation is calculated by:
\begin{equation}
Q_{wall}= h(T- T_{amb}) + \epsilon\sigma(T^4 -T_{amb}^4 )
\label{eq:turbulent_conductivity}
\end{equation}
\RVt{where \(T_{\mathrm{amb}}\) represents the room temperature and is fixed at 300~K, and \(h\) is an effective heat-transfer coefficient set to 25~W~m$^{-2}$~K$^{-1}$, following the value used by Wolf \textit{et al.} \cite{wolf2020co2}. This boundary condition represents heat exchange between the outer tube wall and the surrounding air, but does not imply active air cooling of the discharge tube.} \(\epsilon\) is the emissivity of quartz fixed at 0.93, and \(\sigma\) is the Stefan-Boltzmann constant \cite{wolf2020co2}.

Meshing refers to the discretization of a computational geometry into a finite number of smaller elements, forming a numerical grid for solving the governing equations. The resulting mesh consists of nodes, faces, and cells, which may take various shapes, such as tetrahedral, hexahedral, or poly-Hexcore elements \cite{jeong2014comparison}. Table \ref{tab:placeholder_label} summarizes the influence of different meshing strategies on the key simulation parameters under identical mesh size conditions. The orthogonal quality is a mesh quality metric that quantifies the degree to which a computational cell deviates from an ideal orthogonal configuration. The value ranges from 0 to 1, where values closer to 1 indicate higher mesh quality, improved numerical stability, and reduced discretization errors. Due to its high level of automation, the tetrahedral mesh is widely used for complex geometries. However, tetrahedral meshes generally suffer from lower numerical accuracy and require a significantly larger number of cells, which leads to increased computational cost and memory consumption \cite{biswas1998tetrahedral}. The poly-Hexcore mesh, as recommended by ANSYS Fluent, combines high-quality octree-based hexahedral cells in the core region with isotropic polyprism elements near the boundaries, connected through Mosaic polyhedral elements \cite{matsson2022introduction}. This hybrid meshing strategy provides improved mesh quality, particularly in terms of cell orthogonality and reduced numerical diffusion \cite{zore2019ansys}. Compared with a purely tetrahedral mesh, the poly-Hexcore mesh typically requires only about 22\% of the computational cost, while delivering superior accuracy and numerical stability. Therefore, the poly-Hexcore mesh is adopted in the present work.

\begin{table}[ht]
    \centering
    \caption{Comparison of mesh parameters using the same reactor geometry and mesh size.}
    \label{tab:placeholder_label}
    \begin{tabular}{lcc}
        \toprule
        Method & Orthogonal Quality &  Mesh cells number \\
        \midrule
        Tetrahedral & 0.15198901 & 5,587,001 \\
        Hexcore     & 0.15614507 & 1,728,425 \\
        Poly-hexcore & 0.20781557 & 1,239,173 \\
        \bottomrule
    \end{tabular}
\end{table}

Furthermore, mesh resolution also plays a crucial role in CFD simulations, as it directly affects the accuracy, convergence behavior, and computational efficiency of the numerical solution. A finer mesh can resolve flow features such as boundary layers and temperature gradients more accurately, improving the spatial resolution of the predicted fields.  However, it also leads to a substantial increase in the number of computational cells, resulting in higher memory requirements and longer computational times. Consequently, mesh optimization becomes a key strategy for balancing numerical accuracy and computational efficiency.  
\RVt{To reduce the computational cost while retaining sufficient resolution in the most reactive region, the reactor domain was divided into three refinement zones. The inner region covers the emission-based plasma zone, where the steepest temperature gradients, strongest chemical source terms, and largest species gradients are expected. The medium region surrounds the plasma and resolves the transport between the hot plasma core and the colder outer flow, including recirculation, shear, heat transfer, and species diffusion near the plasma periphery. The outer region covers the remaining relatively cold-flow region and the downstream reactor volume, where the gradients are generally weaker and a coarser mesh can be used without strongly affecting the main temperature and conversion results. A schematic showing the inner, medium, and outer mesh-refinement regions is provided in the Supporting Information.}

\begin{table}[h]
\centering
\caption{\RVt{Mesh-independence test cases, target mesh sizes, calculation times, and memory requirements.}}
\label{tab:mesh_independence}
\begin{tabular*}{\textwidth}{@{\extracolsep{\fill}}ccccccc}
\toprule
Case & Elements & Inner [mm] & Medium [mm] & Outer [mm] & Time & Memory [GB] \\
\midrule
1 & 240,477  & 0.30 & 0.90 & 1.80 & 3:03:18  & 11.9 \\
2 & 545,380  & 0.25 & 0.75 & 1.50 & 4:37:31  & 13.2 \\
3 & 1,239,173 & 0.20 & 0.60 & 1.20 & 13:20:00 & 16.3 \\
4 & 3,023,280 & 0.15 & 0.45 & 0.90 & 36:06:27 & 24.6 \\
\bottomrule
\end{tabular*}
\end{table}

Table \ref{tab:mesh_independence} summarizes the four mesh cases tested in this work, including the target mesh sizes in the inner, medium, and outer refinement regions. All calculations were performed on the same workstation equipped with a 13th Gen Intel(R) Core(TM) i7-13700K processor (16 cores, 24 logical processors) and 96~GB RAM. The calculation time increased from 3~h~03~min for Case~1 to 36~h~06~min for Case~4, while the memory requirement increased from 11.9 to 24.6~GB. Figure~\ref{fig:T_different_mesh} compares the axial temperature profiles obtained for different cases with varying mesh resolutions. The third case yields temperature profiles nearly identical to those of the finest mesh. Furthermore, the difference in conversion results between these two cases is below 1\%, indicating mesh-independent behavior. Consequently, the third case is employed in all subsequent simulations, offering a good compromise between accuracy and computational efficiency.

\begin{figure}[ht]
\centering
\includegraphics[width=0.6\linewidth]{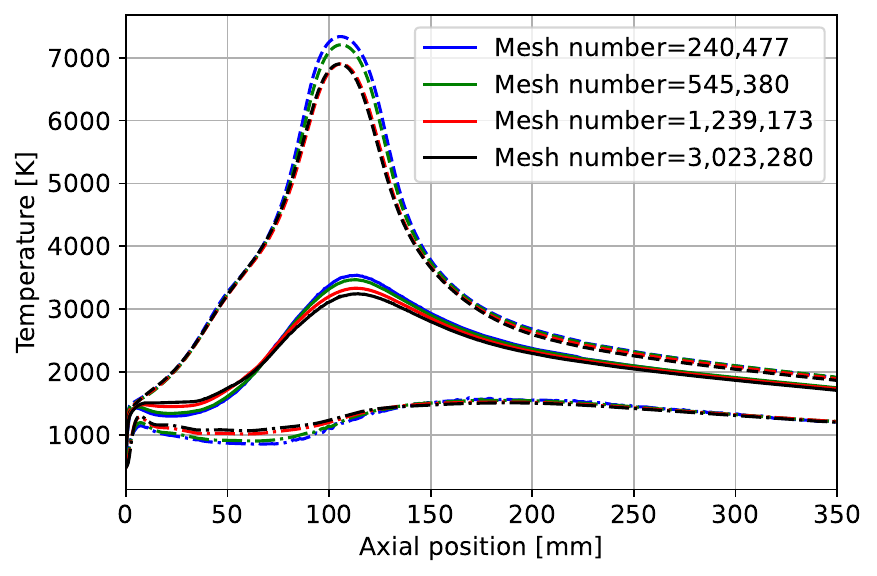}
\caption{Axial temperature profiles at different radial positions for various mesh resolutions under the 150~mbar condition. The dashed, solid, and dash-dotted lines correspond to radial positions of 0, 5, and 10~mm, respectively.}
\label{fig:T_different_mesh}
\end{figure}
}

\subsection{Chemistry kinetics}
\RVt{The three mechanisms considered in this work are the Park, Johnston, and GRI-Mech-based mechanisms. The Park mechanism was originally developed for high-temperature thermochemical non-equilibrium in atmospheric-entry flows \cite{park1989nonequilibrium}. The Johnston mechanism is based on the Park kinetic framework and modifies selected reaction rates to improve the prediction of non-equilibrium CO and CN radiation \cite{johnston2014modeling}. The most notable modification concerns the CO dissociation rate, which was increased relative to the previously adopted value.} \RVt{The third mechanism is based on GRI-Mech, which was originally developed for high-temperature gas-phase combustion and has been widely used in thermochemical reaction modeling \cite{gri-mech}. Because the original GRI-Mech mechanism does not explicitly include CO dissociation, the rate coefficient for this reaction is supplemented from Johnston \textit{et al.} \cite{johnston2014modeling} and added to the GRI-Mech-based reaction set.} \RVt{All three mechanisms are implemented for the five species included in the present model, including CO$_2$, CO, C, O, and O$_2$. Their differences in rate coefficients provide a basis for evaluating the sensitivity of the predicted temperature field and CO$_2$ conversion to the selected chemical kinetics. }

\RV{Furthermore,  ozone formation through the three-body reaction may become more favorable after the gas has cooled, particularly at elevated pressure. However, O$_3$ is not expected to \RVt{accumulate} significantly in the hot plasma region and early \RVt{afterglow}, where most CO$_2$ conversion occurs. Therefore, its omission is expected to have a limited influence on the predicted CO yield, although it may affect the downstream distribution of O and O$_2$. The present mechanism cannot quantify this effect, and an extended mechanism including ozone and electronically excited species should be considered in future work.} 

\RV{Catalytic wall recombination of O atoms is not included in the present model. This treatment is consistent with the radial fluid model of Vialetto \textit{et al.} \cite{vialetto2022charged}, where wall fluxes were included but surface recombination of atomic species was neglected. They reported that, under comparable microwave CO$_2$ plasma conditions, the species composition is mainly controlled by gas-phase chemistry rather than surface reactions. Consistently, the present results show that O atoms are strongly confined to the hot plasma core and decrease sharply toward the wall owing to the pronounced radial temperature gradient. Therefore, only a limited O-atom flux is expected to reach the quartz surface, and wall recombination is assumed to have a minor influence on the overall O-atom balance and CO yield. Nevertheless, material-dependent wall-recombination probabilities should be considered in future work to quantify the associated uncertainty.}

The net rate of reaction \(j\) is defined as the difference between the forward and reverse rates:
\begin{equation}
R_j = R_j^f- R_j^r
    = k_j^f \prod_l n_l^{\nu'_{lj}}
    - k_j^r \prod_l n_l^{\nu''_{lj}},
\label{eq:reaction_Rate}
\end{equation}
where \(R_j^f\) and \(R_j^r\) are the forward and reverse reaction rates of reaction \(j\), respectively. \(k_j^f\) and \(k_j^r\) are the corresponding rate constants. \(n_l\) [mol m$^{-3}$] is the molar density of species \(l\), and \(\nu'_{lj}\) and \(\nu''_{lj}\) are the stoichiometric coefficients of species \(l\) on the reactant side and product side of reaction \(j\), respectively.

\section{Results and discussion}

\subsection{Pressure-dependent plasma behavior and thermochemical model validation}

\begin{figure}[h]
\centering
\includegraphics[width=1\linewidth]{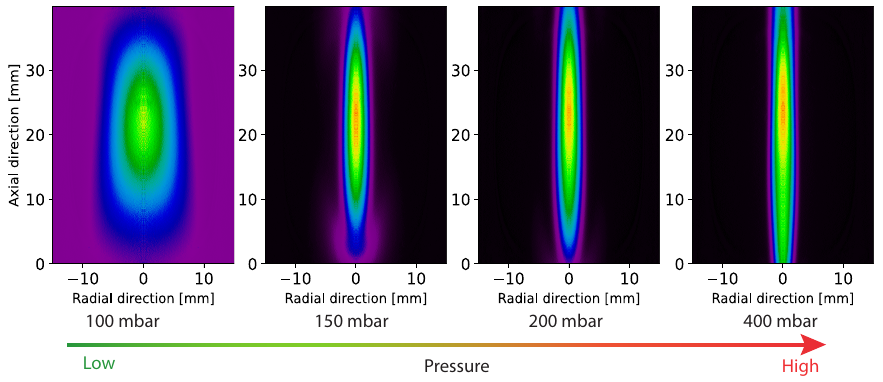}
\caption{Recorded profiles of O 777 nm emission by CCD detector at different pressures \Lex{after Abel inversion}. \RVt{The colour scale represents relative emission intensity in the experiments only; it should not be interpreted as a velocity or recirculation map.}}
\label{fig:plasma_picture}
\end{figure}

\RVt{
Figure~\ref{fig:plasma_picture} and Tab.~\ref{tab:experimental_conditions} show that increasing pressure leads to a strong radial contraction of the plasma, together with an axial elongation of the emission region. The plasma radius decreases from \(5.5\pm0.3\) mm at 100 mbar to \(1.3\pm0.1\) mm at 400 mbar, whereas the plasma length increases from \(19.6\pm2.5\) mm to \(41.7\pm5.4\) mm. The estimated plasma volume, decreases sharply from 100 to 200 mbar because of the radial contraction, and then increases slightly from 200 to 400 mbar as the plasma length continues to grow while the radius remains almost constant. } \RVt{The pressure-dependent contraction of CO$_2$ microwave plasmas has been discussed in more detail in our previous studies \cite{wolf2019characterization, wolf2020implications}.}

\begin{figure}[ht]
\centering
\includegraphics[width=0.5\linewidth]{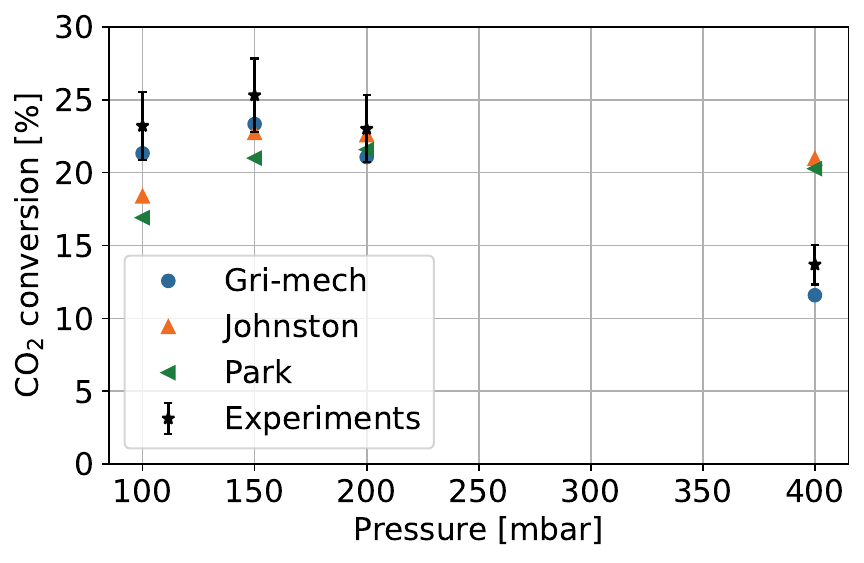}
\caption{Comparison of experimentally measured CO$_2$ conversion with model predictions obtained using different chemical mechanisms at different pressures.}
\label{fig:conversion}
\end{figure}

Figure~\ref{fig:conversion} compares the experimentally measured CO$_2$ conversion with model predictions obtained using different chemical mechanisms as a function of pressure. The experiments exhibit a clear \RVt{non-monotonic} pressure dependence, with the CO$_2$ conversion reaching a maximum at 150~mbar and decreasing at higher pressures, \RVt{consistent with previous experimental observations in microwave CO$_2$ plasmas \cite{van2024effluent, wolf2019characterization}. Briefly, at low pressure, the gas velocity is higher, and the \RVt{heavy-particle} collision frequency is lower, which limits CO$_2$ dissociation despite the relatively broad plasma region. At intermediate pressure, higher gas temperature and sufficient reactive volume favor CO formation. At higher pressure, the plasma contracts radially and three-body recombination becomes stronger, which promotes CO oxidation back to CO$_2$ and lowers the net conversion. This pressure-dependent balance between CO formation and recombination is discussed in detail in the next section. The agreement at lower pressures is particularly noteworthy, as it shows that the measured conversion can be reproduced by a heavy-particle thermochemical CFD model within the present pressure range. This is consistent with previous diagnostics and kinetic modeling, which reported limited vibrational non-equilibrium and a minor direct contribution of electron-impact CO$_2$ dissociation under comparable conditions \cite{van2021redefining,vialetto2022charged, viegas2020insight}. }

\begin{table}[ht]
    \centering
    \caption{Experimentally determined plasma geometric parameters and CO$_2$ conversion rates at different pressures. The plasma volume is estimated by approximating the emission envelope as a cone.}
    \label{tab:experimental_conditions}
    \begin{tabular}{cccccc}
        \toprule
        Pressure [mbar] & \(\alpha\) [\%] & \(\eta\) [\%] & \(R_{\mathrm{plasma}}\) [mm] & \(H_{\mathrm{plasma}}\) [mm] & \(V_{\mathrm{plasma}}\) [mm$^3$] \\
        \midrule
        100 & 23.3\(\pm\)2.3 & 32.5\(\pm\)3.2 & 5.5\(\pm\)0.3 & 19.6\(\pm\)2.5 & 621\(\pm\)104 \\
        150 & 25.3\(\pm\)2.5 & 35.3\(\pm\)3.5 & 2.1\(\pm\)0.1 & 25.3\(\pm\)3.3 & 117\(\pm\)19 \\
        200 & 23.0\(\pm\)2.3 & 32.1\(\pm\)3.2 & 1.4\(\pm\)0.1 & 31.0\(\pm\)4.0 & 64\(\pm\)12 \\
        400 & 13.7\(\pm\)1.4 & 19.1\(\pm\)1.9 & 1.3\(\pm\)0.1 & 41.7\(\pm\)5.4 & 74\(\pm\)15 \\
        \bottomrule
    \end{tabular}
\end{table}

Among the three chemical mechanisms considered, the GRI-Mech-based model shows \RVt{the best} agreement with the experimental data across the entire pressure range. In particular, it \RVt{reproduces} both the peak CO$_2$ conversion at 150~mbar and the pronounced decrease \RVt{at} 400~mbar. This agreement suggests that GRI-Mech captures the \RVt{main} pressure-dependent reaction pathways \RVt{controlling} CO formation and consumption. In contrast, the Johnston and Park mechanisms \RVt{deviate} from the experimental results at both low and high pressures. These discrepancies indicate that, although the overall pressure trend is reproduced, important kinetic limitations and recombination effects at the pressure extremes are not fully captured by these mechanisms. A plausible \RVt{reason for the better performance of the GRI-Mech-based mechanism is} its pressure-dependent fall-off treatment of the CO$_2$ dissociation reaction (reaction N$^r_1$), \RVt{which is shown in Tab.~\ref{tab:reactions}}. It gives different effective reaction rates under \RVt{different} pressure conditions.

\RVt{The temperature fields obtained with the different kinetic mechanisms show that the core temperature varies by only a few hundred kelvin, depending on pressure, illustrating that the temperature field is not fully independent of the kinetic scheme. The difference comes from the mechanism-dependent balance between endothermic dissociation, recombination, and heat transport in the high-temperature region. Since the GRI-Mech-based mechanism gives the closest agreement with the measured conversion trend and radial temperature profiles, it was used for the following analysis.} \RVt{The corresponding energy-efficiency comparison is provided separately in the Supporting Information, because the efficiency and conversion uncertainties are different and are easier to read in separate panels.}

\begin{table}[t]
\caption{List of GRI--Mech--based chemistry reactions. The rate coefficient of the opposite direction is calculated from the principle of detailed balance using the equilibrium constant \cite{shen2025pinpointing}. }
\centering
\label{tab:reactions}
\begin{tabular}{cccc}
 \hline
 No. & Reaction & Rate coefficient  & Ref    \\
 \hline

  & &    \(k_o=6.02\times\)10$^{14}$ \text{exp}(-12552/R\textit{T}) [cm$^6$(mol$^2$s)$^{-1}$]       &  \\
 {$^a$}N$_1$ & \(\text{CO$_2$}  + \text{M} \leftrightarrow \text{CO} + \text{O} + \text{M}\)          &   \(k_{inf}=1.8\times\)10$^{10}$ \text{exp}(-9978.8/R\textit{T}) [cm$^3$(mol s)$^{-1}$]&  \cite{gri-mech}\\

 &     &   \(k_{1}^{r}=  k_0/(1+k_0/k_{inf})  \)    &\\
 \hline
 {$^b$}N$_2$ & \(\text{CO}  + \text{M} \leftrightarrow   \text{C}+ \text{O} + \text{M}\) &    \(k_2^f=1.2\times\)10$^{21}$\textit{T}$^{-1}$ \text{exp}(-61268.8/R\textit{T}) [cm$^3$(mol s)$^{-1}$] &   \cite{johnston2014modeling}\\
 \hline

 {$^c$}N$_3$ & \(\text{O$_2$} + \text{M} \leftrightarrow \text{O} + \text{O}   + \text{M}\) &    \( k_3^r=1.2\times\)10$^{17}$\textit{T}$^{-1}$ [cm$^6$(mol$^2$s)$^{-1}$] & \cite{gri-mech} \\
 \hline

 N$_4$ & \(\text{CO$_2$} + \text{O} \leftrightarrow \text{CO} + \text{O$_2$}\) &    \(k_4^r=2.5\times\)10$^{12}$ \text{exp}(-11424.5/R\textit{T}) [cm$^3$(mol s)$^{-1}$] & \cite{gri-mech} \\
 \hline

 N$_5$ & \(\text{C} + \text{O$_2$}   \leftrightarrow \text{CO} + \text{O}\) &    \(k_5^r=5.8\times\)10$^{13}$ \text{exp}(-137.7/R\textit{T}) [cm$^3$(mol s)$^{-1}$] & \cite{gri-mech} \\
 \hline
\end{tabular}
\footnotesize{{$^a$} Rate  coefficient is multiplied by 1.5/6.0/3.5 for M = CO/O$_2$/CO$_2$.}

\footnotesize{{$^b$} Rate  coefficient is multiplied by 1.75 for M = CO.}

\footnotesize{{$^c$} Rate  coefficient is multiplied by 1.5 for M = atoms.}

\end{table}

\begin{figure}[ht]
\centering
\includegraphics[width=0.5\linewidth]{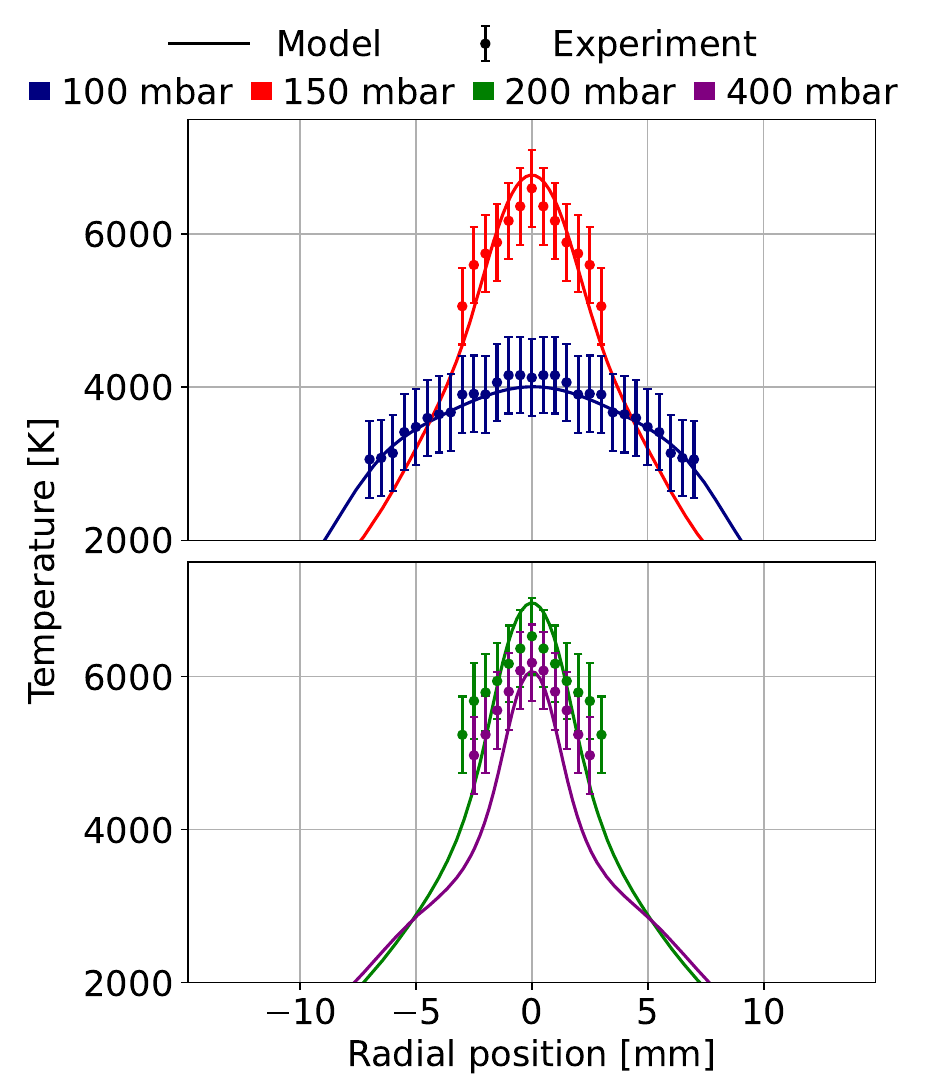}
\caption{Comparison between modeled and experimentally measured radial temperature profiles at the plasma center under different pressure conditions.}
\label{fig:T_radial_model_ex}
\end{figure}

In addition to overall conversion validation, Figure~\ref{fig:T_radial_model_ex} compares the experimentally measured and numerically \RVt{simulated} radial gas temperature profiles in the plasma core at pressures of 100–400~mbar. The model accurately reproduces both the radial temperature distributions and their pressure-dependent evolution. At 100~mbar, both experiments and simulations show a broad, smooth profile with peak temperatures of 3500–4000~K. \RVt{With pressure increasing, the discharge contracts radially and the core gas temperature rises rapidly to 6000--7000~K.  The increase in core temperature is mainly associated with the sharp reduction in plasma volume (Tab.~\ref{tab:experimental_conditions}), which concentrates the deposited power into a more confined discharge region. However, the model underestimates several off-axis temperatures, especially at higher pressures. One possible contribution is the larger uncertainty of Doppler-broadening thermometry near the plasma edges. Under these conditions, the discharge is less stable, and the emission boundary may shift during the measurement. Because the temperature is obtained from line-of-sight O 777 nm emission and spectral profile fitting, this motion can broaden the effective sampling region and increase the uncertainty of the off-axis values. The prescribed heat-source profile may also add to this difference, since it cannot capture weak off-axis power deposition or transient displacement of the plasma edges. We therefore treat the off-axis mismatch as a limitation of the present validation, while noting that the model reproduces the main radial temperature trend and the pressure-dependent conversion behavior.}

\begin{figure}[h]
\centering
\includegraphics[width=0.9\linewidth]{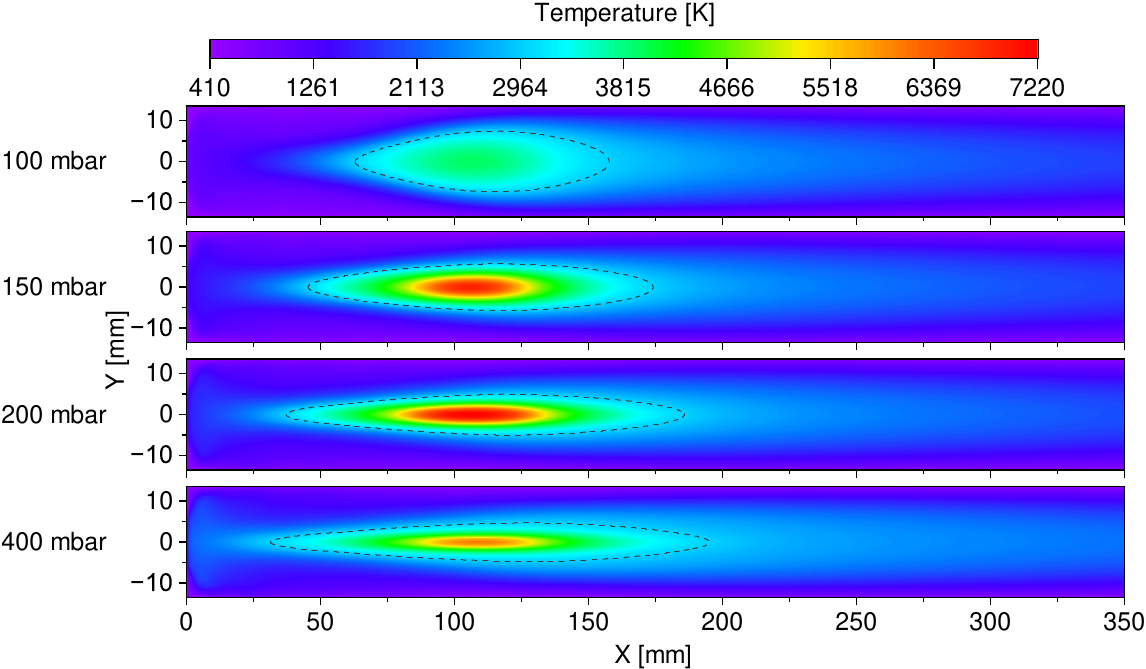}
\caption{Temperature distributions in a cross-sectional plane at the mid-height of the reactor under different pressure conditions. \Lex{Only the axial region from 0 to 350 mm is shown. The dashed line denotes the boundary of 3000 K.} }
\label{fig:T_different_pressure}
\end{figure}

\RVt{Based on this validation, the simulated temperature fields are then used to assess how pressure changes the thermal structure of the full reactor.} Figure~\ref{fig:T_different_pressure} shows that the high-temperature region evolves from a broad, short plasma column at 100~mbar to a radially contracted and axially elongated structure at higher pressure. \RVt{At 100~mbar, the temperature field remains characteristic of a diffusion discharge. Previous Thomson-scattering measurements and kinetic modeling for vortex-stabilized CO$_2$ microwave discharges reported electron temperatures of the order of 1--2~eV in this pressure range, above the corresponding gas temperature \cite{van2022chemical}. However, the ionization degree remains low. Viegas \textit{et al.} reported ionization degrees of $0.76\times10^{-5}$ at 97~mbar and $1.85\times10^{-5}$ at 108~mbar for a similar CO$_2$ microwave discharge \cite{viegas2020insight}. Thus, the higher electron temperature does not imply a large electron thermal-energy reservoir. Because the electron density is about five orders of magnitude lower than the heavy-particle density, most of the absorbed microwave power can still be transferred to gas heating and heavy-particle chemistry. This explains why the present thermochemical CFD model can reproduce the measured gas-temperature profile even at 100~mbar when the experimentally measured heat-source profile is imposed.} \RVt{The temperature field is also shaped by endothermic chemistry.} The 3500--4000~K range corresponds to the onset of substantial CO$_2$ dissociation \cite{snoeckx2017plasma}, so part of the deposited energy is consumed by CO$_2$ dissociation rather than by sensible gas heating. At 150 and 200~mbar, the contracted plasma volume increases the local deposited power density, and the high-temperature region extends further downstream. In this temperature range, CO$_2$ dissociation becomes nearly complete and CO dissociation also becomes relevant. At 400~mbar, the discharge enters the high-confinement regime. Compared with 200~mbar, the plasma length increases while the radius changes only slightly, reducing the volumetric power density (Tab.~\ref{tab:experimental_conditions}). The core gas temperature at 400~mbar is therefore slightly lower than at 200~mbar.

Overall, the proposed thermal–chemical-flow model shows good agreement with experimental observations from 100~mbar to 400~mbar. The model accurately captures both the pressure-dependent evolution of the radial gas temperature profiles, as well as the non-monotonic variation of CO$_2$ conversion with pressure. When combined with the GRI-Mech mechanism, the model provides the closest quantitative agreement with the CO$_2$ conversion measurements among the mechanisms considered. These results indicate that the model reliably describes the dominant thermochemical processes and pressure-dependent reaction kinetics in the MW CO$_2$ plasma reactor, thereby supporting its suitability for further mechanistic analysis.

\subsection{Pressure-dependent CO formation and reaction pathways}

\subsubsection{Species distributions and reaction pathways at 150~mbar}

\begin{figure}[h]
\centering
\includegraphics[width=0.8\linewidth]{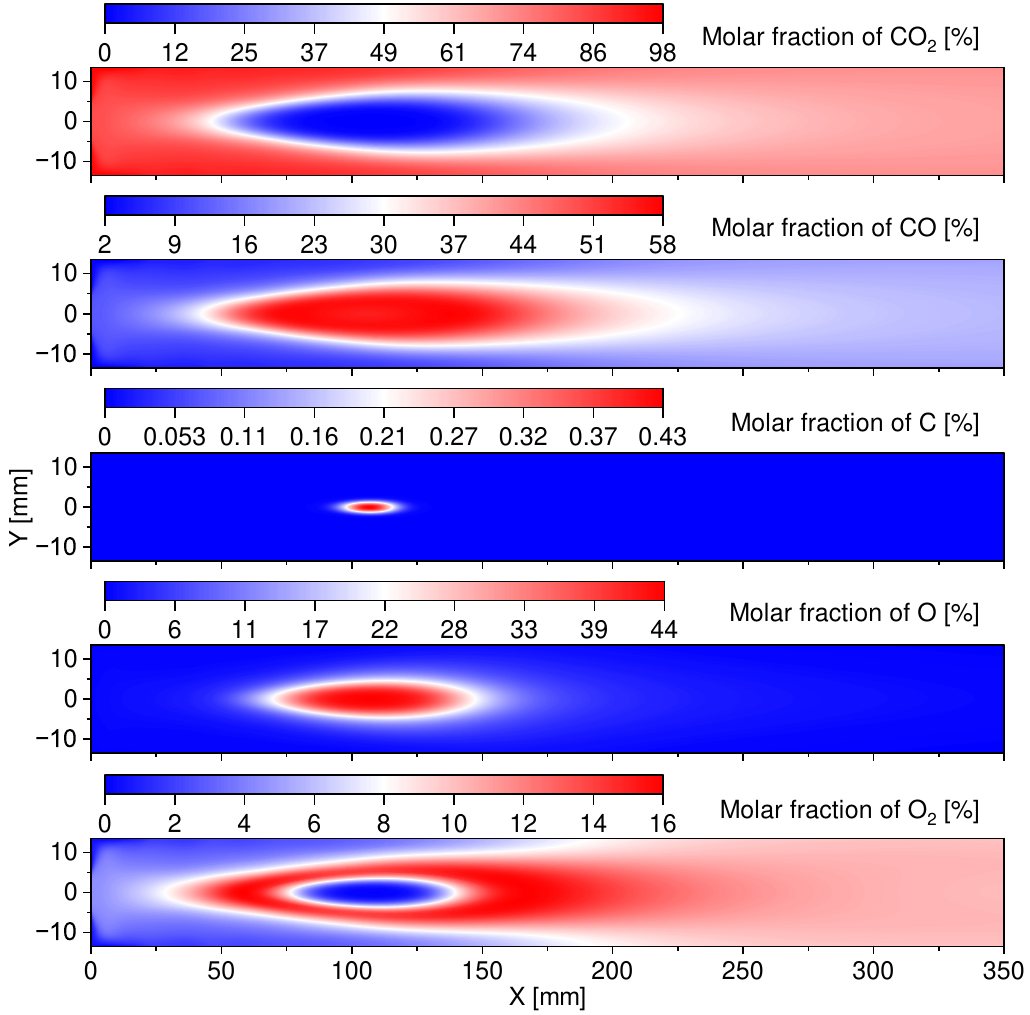}
\caption{Species distributions in a cross-sectional plane at the mid-height of the reactor under 150 mbar pressure conditions. \Lex{Only the axial region from 0 to 350 mm is shown.}  }
\label{fig:molar_fraction_150mbar}
\end{figure}

\begin{figure}[h]
\centering
\includegraphics[width=0.75\linewidth]{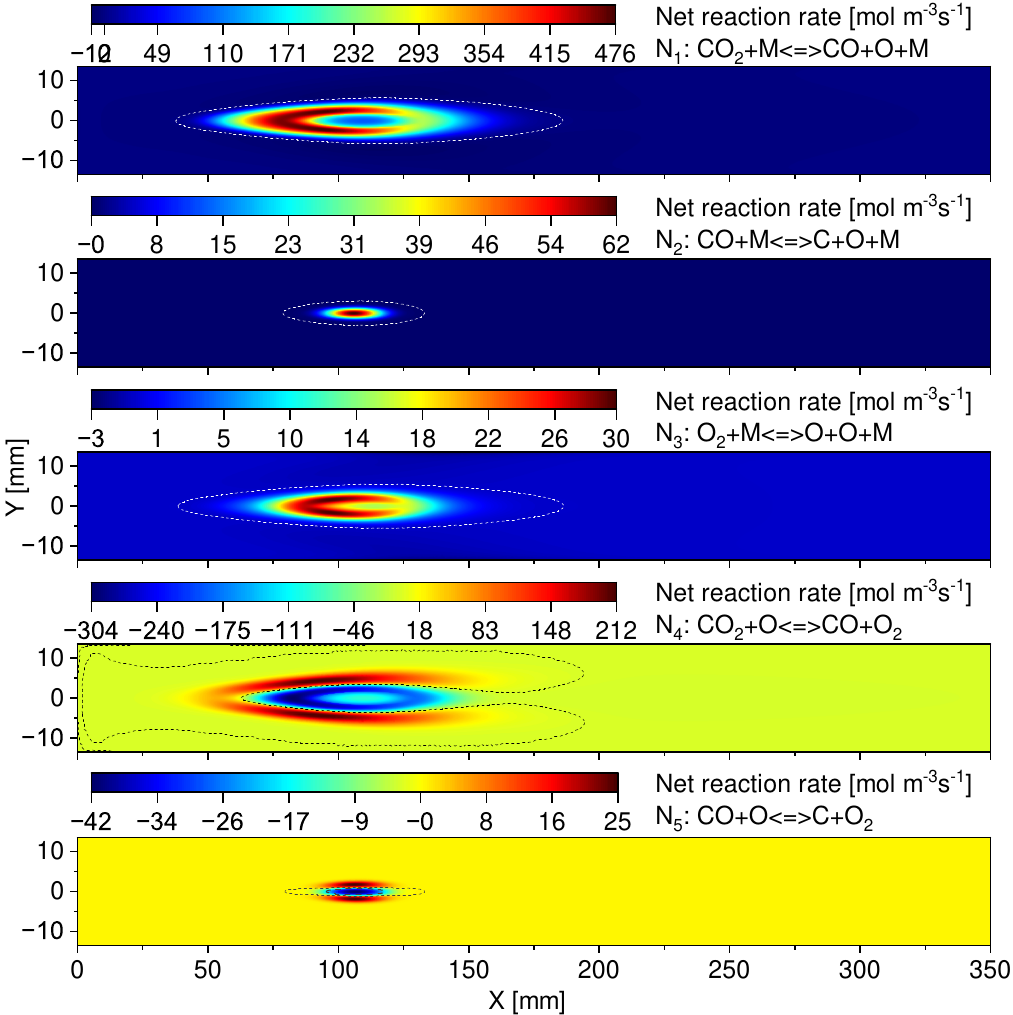}
\caption{Net reaction rate distributions in a cross-sectional plane at the mid-height of the reactor under 150 mbar pressure conditions. \Lex{Only the axial region from 0 to 350 mm is shown.} The dashed line denotes the boundary separating positive and negative net reaction rates.}
\label{fig:net_rate_150mbar}
\end{figure}

\RVt{Since the highest CO$_2$ conversion is obtained at 150~mbar, this condition is analyzed in detail to identify the local reaction environment responsible for efficient CO formation. The species mole-fraction fields (Fig.~\ref{fig:molar_fraction_150mbar}) and net reaction-rate distributions (Fig.~\ref{fig:net_rate_150mbar}) are considered together in this subsection, as the dominant pathways are governed by the spatial overlap between temperature, reactant availability, and transport from the surrounding CO$_2$-rich region.} As shown in Fig.~\ref{fig:molar_fraction_150mbar}, CO and O are concentrated predominantly within the plasma core, whereas O$_2$ is formed mainly near or outside its edges. It has good agreement with Raman-scattering measurements of van de Steeg \textit{et al.} for a comparable microwave CO$_2$ plasma at 120~mbar \cite{van2021redefining}. \RVt{Although the model is quantitatively validated against the radial gas-temperature distribution and the overall CO$_2$ conversion, future spatially resolved Raman-scattering measurements are required to validate the individual species distributions more rigorously.}

\RVt{Reaction-rate analysis shows that reactions N$_1$ and N$_4$ (Tab.~\ref{tab:reactions}) dominate the local CO$_2$ conversion rates at 150~mbar. In the plasma core, these two reactions have opposite effects. The reverse reaction N$_4$ reforms part of the CO$_2$ from CO and O$_2$, whereas reaction N$_1$ consumes CO$_2$. The CO$_2$ available for reaction N$_1$ is maintained by local reformation through reverse reaction N$_4$ and by transport from the surrounding CO$_2$-rich region. As a result, around 3.5\% CO$_2$ mole fraction remains in the plasma core.} \RVt{At the plasma edges and in the upstream region, reaction N$_4$ becomes the dominant local CO-production pathway. This allows additional CO$_2$ conversion at about 2000--3000~K and extends the effective reaction zone beyond the hottest plasma core.}

\RVt{Although the local reaction rates of N$_1$ and N$_4$ (Tab.~\ref{tab:reactions}) are negligible in the reactor-top region ($x=0$--25~mm), relatively high CO molar fractions of about 10--20\% persist there.} It is primarily attributed to flow-induced particle redistribution, whereby CO generated in the plasma core is transported upstream via recirculation. It will be discussed in detail in the next section. This highlights the critical role of flow structures in shaping species distributions.

\RVt{Atomic oxygen reaches its highest mole fraction in the plasma core and decreases rapidly in both the radial and axial directions, indicating that it is produced mainly in the high-temperature dissociation region. The reaction-rate analysis identifies reaction N$_1$ and the reverse reaction N$_4$ as the main local sources of O atoms, whereas reaction N$_3$ and the reverse reaction N$_5$ make only minor contributions. The O-atom mole fraction remains very low near the reactor wall because of the steep radial temperature gradient and the strong localization of O production in the plasma core. This supports the assumption that catalytic wall recombination of O atoms has only a limited influence on the reactor-scale O balance and CO yield under the present conditions.}

In the \RVt{afterglow}, overall reaction activity is substantially reduced. The rates of reactions N$_1$ and N$_4$ become negative, indicating that these reactions contribute to CO loss. Furthermore,  diffusion plays a more vital role in the transport mechanism, progressively homogenizing the radial species distributions and marking a transition from a reaction-controlled to a transport-dominated regime. These results demonstrate that the spatial distributions of species are governed not solely by local reaction kinetics, but also by transport processes \cite{wolf2020co2}.

\subsubsection{Pressure-dependent CO production and reactor-integrated reaction contribution}

\begin{figure}[h]
\centering
\includegraphics[width=0.9\linewidth]{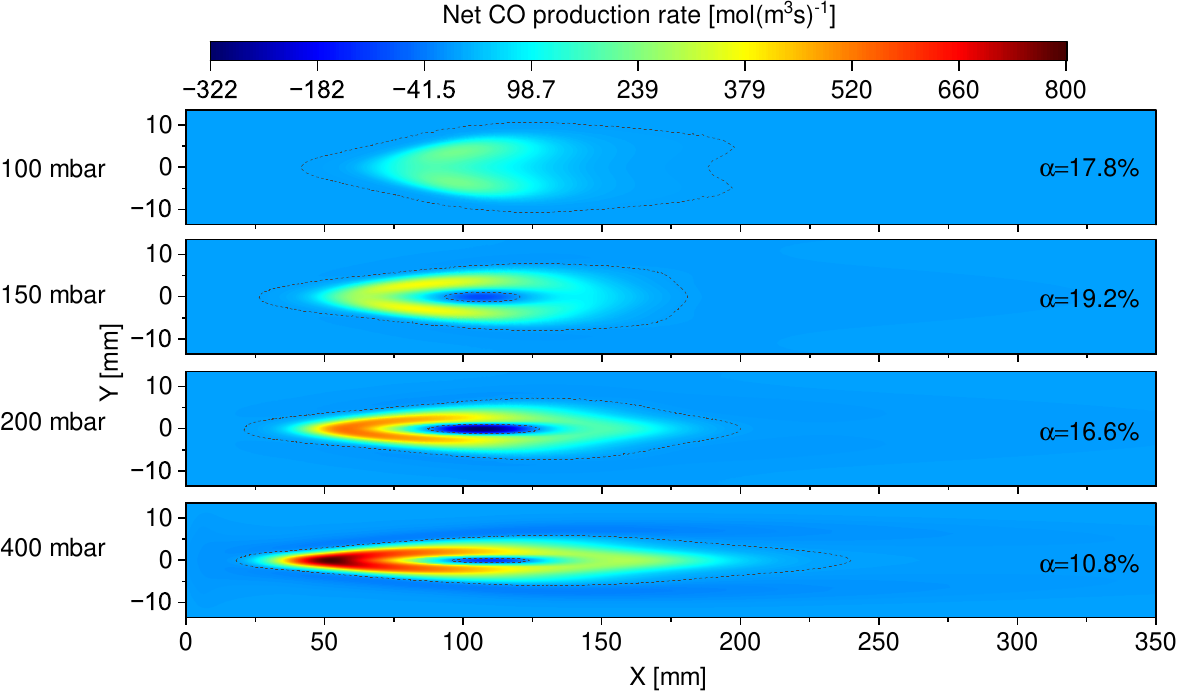}
\caption{\Lex{Net CO production rate} distributions in a cross-sectional plane at the mid-height of the reactor under different pressure conditions. \Lex{Only the axial region from 0 to 350 mm is shown.} The dashed line denotes the boundary separating positive and negative net reaction rates.  }
\label{fig:net_CO_production_different_pressure}
\end{figure}

Across all pressures, net CO production shows a highly localized and non-uniform pattern that is closely linked to the gas temperature, flow structure, and active plasma region (Fig.~\ref{fig:net_CO_production_different_pressure}). At 100~mbar, the overall net CO production rate remains low, with positive values confined mainly to the upstream edges of the plasma core and without pronounced axial or radial extension. \RVt{Although the gas temperature inside the plasma reaches approximately 3000--4000~K, which lies in the favorable temperature range for CO$_2$ dissociation, the lower heavy-particle collision frequency and the higher gas velocity keep CO$_2$ dissociation in a kinetically limited regime. The velocity magnitude at different pressures can be found in the Supporting Information. Notably, no net CO consumption is observed in the plasma region, indicating that secondary CO-consuming pathways are not dominant under this condition.}

\RVt{As the pressure is increased to 150~mbar, the peak net CO production rate increases by nearly a factor of two compared with the 100~mbar case and exhibits a more complex spatial structure. On the one hand, the region with positive net CO production also extends upstream by approximately 15~mm.} On the other hand, localized regions of negative net CO production emerge in the plasma core. \Lex{This transition reflects the fact that, although CO$_2$ dissociation is substantially enhanced with increasing pressure, reverse reactions leading to CO$_2$ reformation are simultaneously activated (Fig.~\ref{fig:net_rate_150mbar}).  CO$_2$ dissociation remains dominant near the plasma \RVt{edges}, contributing to positive net CO production, whereas near the plasma core, elevated CO concentration and \RVt{lower gas velocity} enhance reverse recombination pathways (reverse reaction N$_4$), leading to locally negative net CO production rates. Although secondary dissociation of CO into atomic carbon may occur in the hottest regions, this process primarily redistributes chemical energy and does not represent a permanent loss of CO within the present reaction mechanism. Near the downstream end of the plasma, the net CO production becomes positive again, mostly attributed to the entrainment of fresh CO$_2$ into regions that remain sufficiently hot for dissociation. As a result, CO formation and reformation become spatially separated in the plasma region, giving rise to a characteristic axial “positive–negative–positive” structure in the net CO production profile.}

At 200~mbar, the net CO production rate is further enhanced, and the region of positive net CO production extends over a larger axial domain. CO$_2$ dissociation is fully activated \RVt{within the high-temperature region}, enabling efficient CO formation both upstream of the plasma and in the radially outer regions. \RVt{However, this local enhancement does not lead to a higher overall CO$_2$ conversion. The estimated plasma volume decreases from approximately 117~mm$^3$ at 150~mbar to 64~mm$^3$ at 200~mbar, which reduces the gas volume exposed to the most favorable conversion conditions. It explains why the overall conversion decreases despite the stronger local reaction rates.} Near the plasma core, the local CO concentration increases significantly, favoring CO consumption through reverse reaction N$_4$ and \RVt{CO dissociation through reaction N$_2$. The rate profiles of different reactions at 200 mbar can be found in the Supporting Information.} \RVt{The peak rate of reaction N$_2$, corresponding to CO dissociation, increases by approximately a factor of 2.5 compared with the 150~mbar case, although the peak gas temperature differs by less than 200~K. This indicates that the stronger local CO dissociation is not caused by the peak temperature alone, but also by the changed spatial overlap between the high-temperature region and the local CO-rich zone.}

With pressure increasing to 400 mbar, although the region of positive net CO production extends further in the axial direction, it becomes compressed in the radial direction \RVt{(Fig.~\ref{fig:T_different_pressure})}. At high pressures, the substantially increased gas number density greatly enhances the \RVt{heavy-particle} collision frequency for three-body recombination reactions. Although the gas temperature in the plasma \RVt{edges} is lower than in the core, it remains sufficiently high to activate these three-body processes. CO and O atoms transported outward from the plasma core can readily recombine through three-body collisions to reform CO$_2$, resulting in net CO consumption. These results further demonstrate that pressure plays a critical role in controlling the spatial balance between CO formation and destruction by regulating the importance of three-body reaction channels.

\begin{figure}[ht]
\centering
\includegraphics[width=0.5\linewidth]{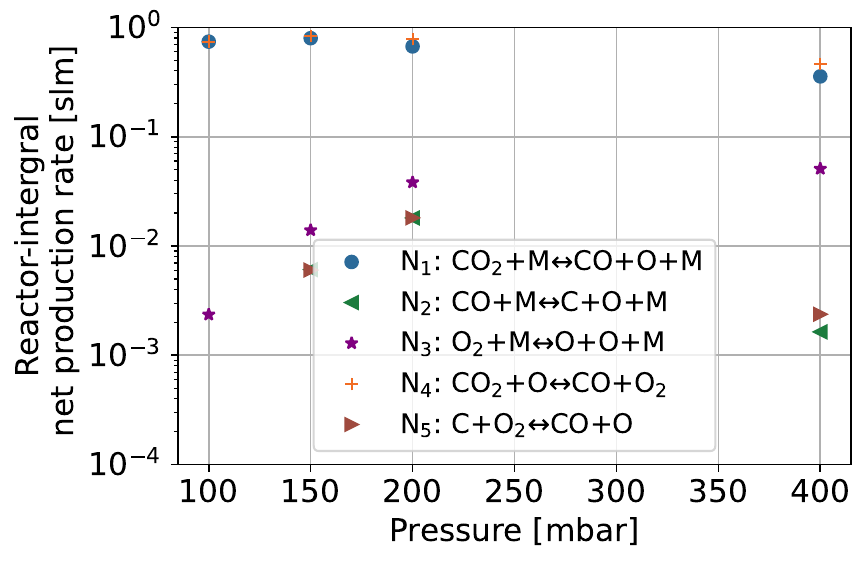}
\caption{Reactor-integrated net contributions of individual reactions throughout the reactor at different pressures. \RVt{Positive values indicate that the forward direction dominates, whereas negative values indicate that the reverse direction dominates.} }
\label{fig:net_reaction_production}
\end{figure}

\RVt{To compare the dominant pathways on a reactor scale, the \RVt{reactor-integrated net contribution $\varphi_j$ [slm]} of reaction \textit{j} is evaluated by integrating its net reaction rate over the entire reactor domain:}
\begin{equation}
\varphi_j=22.414 \times60\times\int R_j dV .
\label{eq:cumulative_production}
\end{equation}
\RVt{where $R_j$ is the net rate of a reversible reaction (Eq.~\ref{eq:reaction_Rate}), including both forward and reverse directions. As shown in Fig.~\ref{fig:net_reaction_production}, reactions N$_1$ and N$_4$ dominate the overall reaction network, with reactor-integrated net production rate on the order of $10^{-1}$~slm across the investigated pressure range. In contrast, the net contributions of reactions N$_2$ and N$_5$ remain several orders of magnitude smaller. At 100 and 150~mbar, reactions N$_1$ and N$_4$ make comparable reactor-integrated net contributions. With increasing pressure, reaction N$_4$ becomes relatively more important, accompanied by a larger net contribution from reaction N$_3$. The additional O atoms produced by reaction N$_3$ favor the O-assisted reaction N$_4$. Meanwhile, higher pressure enhances three-body CO$_2$ recombination, which reduces the net contribution of reaction N$_1$. At 400~mbar, the reactor-integrated net contributions of both reactions N$_1$ and N$_4$ decrease markedly, even though the estimated plasma volume is slightly larger than at 200~mbar (Tab.~\ref{tab:experimental_conditions}). This decrease indicates that the larger plasma volume does not directly translate into higher net CO production, because recombination processes, particularly the pressure-enhanced reverse N$_1$, increasingly offset CO formation in the downstream afterglow. A more detailed discussion will be shown in the last section.}

\subsection{Recirculation structure and turbulent transport}

\subsubsection{Reference flow field at 150~mbar}

\begin{figure}[ht]
\centering
\includegraphics[width=1\linewidth]{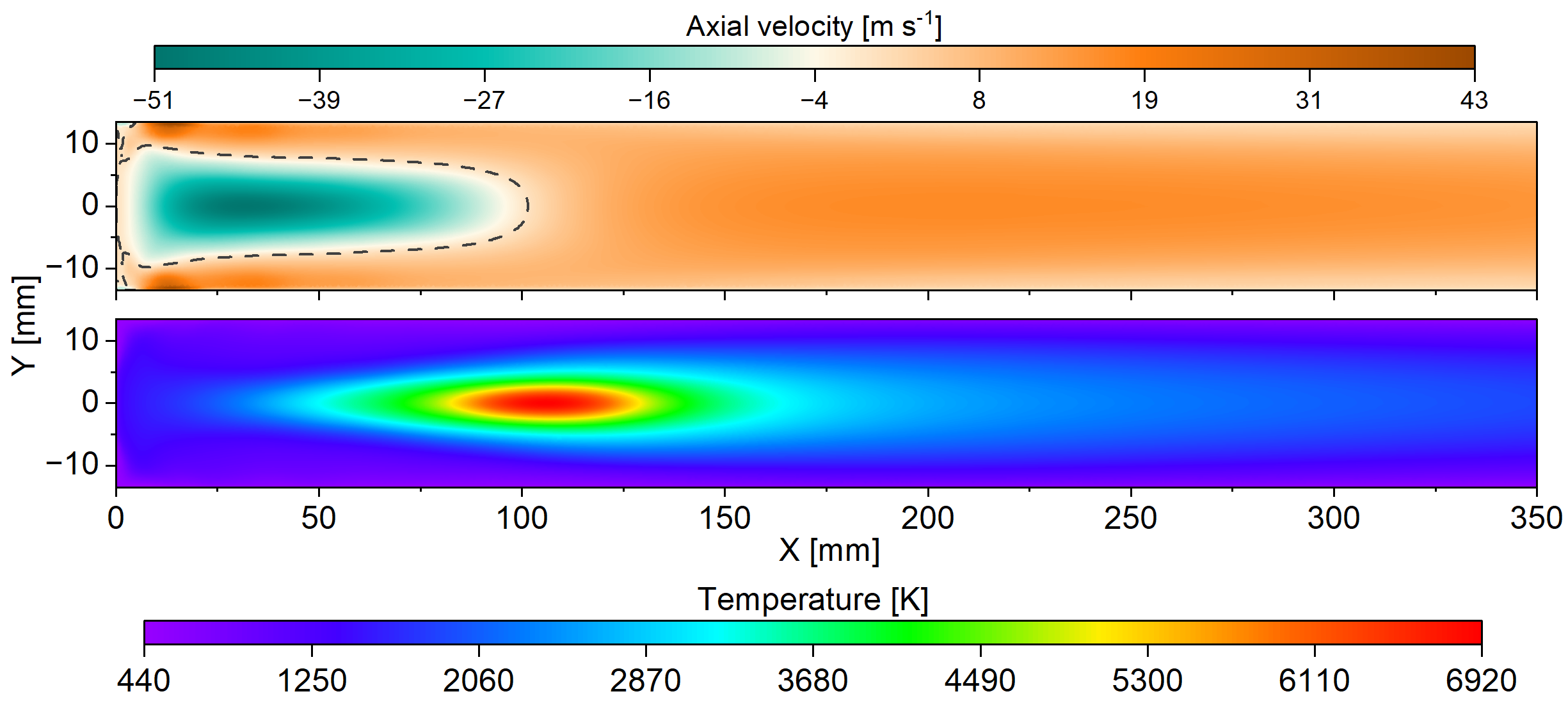}
\caption{Temperature and axial velocity distributions in a cross-sectional plane at the mid-height of the reactor under 150 mbar pressure conditions. The dashed lines indicate the locations where the axial velocity equals zero ($u_x$= 0~m~s$^{-1}$). \Lex{Only the axial region from 0 to 350 mm is shown.}}
\label{fig:v_T_150mbar}
\end{figure}

Gas flow dynamics within a plasma critically determine the performance of chemical processes. However, key parameters, such as velocity distributions, are often difficult to obtain directly through experiments. CFD simulations effectively bridge this experimental gap. \RVt{Here, the 150~mbar case is used as a reference condition for analyzing the flow field, as shown in Fig.~\ref{fig:v_T_150mbar}}. The results reveal a strong axial recirculation structure in the central region of the upstream reactor. It should be noted that the high-temperature zones, though spatially correlated with areas of recirculation and stagnation, are not the dominant driver of the recirculation structure. An isothermal, non-reacting pure fluid model without any external heat source produces a qualitatively similar and even more extensive recirculation pattern, \RVt{as shown in Fig.~S6 in the Supporting Information. It demonstrates that the recirculation structure observed in the present reactor is primarily a hydrodynamic feature inherited from the imposed forward-vortex flow and reactor geometry, rather than a general consequence of plasma formation.} Conversely, the temperature field indirectly influences the flow by modifying gas expansion and temperature-dependent fluid properties. Increased temperature reduces gas density and elevates kinematic viscosity \RVt{(\(\mu/\rho\))}, promoting a faster axial decay of swirl intensity.

Comparison of the axial velocity and static temperature distributions reveals that the recirculation zone does not overlap with the high-temperature region. The maximum temperature is located near the axial position where the axial \Lex{and radial} velocities vanish (\textit{i.e.}, the stagnation point), because convective heat transport is weakened in regions of low axial \Lex{and radial} velocities. Furthermore, \RVt{the lower local gas velocity increases the time over which the gas interacts with the high-power-deposition region, allowing more microwave energy to be transferred to the gas and leading to the formation of a stable high-temperature zone.} It indicates that the high-temperature core is not solely determined by the spatial distribution of energy input, but is also strongly governed by the underlying flow structure. 

\RV{The location of the temperature maximum is also affected by the prescribed power-deposition profile. In the present model, the heat-source distribution is derived from the experimentally measured plasma-emission profile, whose axial peak position changes only marginally over the investigated pressure range. This observation suggests that the main heating region remains relatively stable with pressure. Nevertheless, because the electromagnetic field and plasma properties are not solved self-consistently, some uncertainty remains in the detailed power-deposition profile and in the exact location of the temperature maximum.}

\RVt{A fully coupled electromagnetic--CFD model would be needed to predict the power deposition and temperature field self-consistently. Such a model, however, remains computationally demanding in three dimensions. One possible reduced-dimensional strategy is to first calculate a simplified 3D flow field, extract axial and tangential velocity profiles at a selected axial plane, and impose these profiles as inlet conditions for a 2D-axisymmetric thermochemical model \cite{van2025influence,laitl2025fluid}. For the present reactor, this reduction needs to be treated carefully. The recirculation region already develops between the tangential inlets and the plasma zone. A reduced 2D model would therefore depend on the selected mapping plane and on the physics included before the mapping. Meanwhile, comparison between the flow-only and the full thermochemical models shows that the flow-only case underestimates the axial velocity associated with recirculation (Fig.~\ref{fig:v_T_150mbar} and S6). This illustrates that gas heating, density variation, viscosity variation, and thermal expansion modify the upstream flow field. Developing and validating such a reduced model is left for future work.}

An additional visualization of the flow direction, generated using the Oriented Line Integral Convolution method \cite{matsson2022introduction}, is shown in Fig~S7 in the Supporting Information. The recirculating flow impinges on the reactor \RVt{top} and subsequently changes its direction, migrating toward the near-wall region, and eventually mixes with the \RVt{inlet} gas. This redirection effectively promotes axial transport of the injected gas. Downstream of the recirculation zone, most gas \second{particles} are redirected toward the reactor center, leading to a rapid increase in the axial velocity in the center region. As the flow further develops, the velocity profile gradually evolves toward a Poiseuille-type distribution, characterized by a parabolic radial profile typical of fully developed laminar pipe flow (Fig.~S8 in the Supporting Information). Notably, velocity direction analysis reveals that gas \RVt{streamlines} enter the recirculation zone almost uniformly along the reactor boundary, rather than from a single location. It indicates that the recirculation zone is better described as a volumetric circulation structure rather than an isolated stagnation or dead zone. \RV{Consequently, gas parcels may enter the plasma and its surrounding high-temperature region from different locations, follow substantially different trajectories, and cross these regions once or multiple times. These 3D transport characteristics are expected to produce a broad residence-time distribution, making it difficult to define a single representative residence time for the recirculation zone. A quantitative characterization of this distribution would require Lagrangian particle tracking or an age-of-fluid analysis and will be considered in future work.}

\subsubsection{Pressure dependence of turbulent effect}

\begin{figure}[h]
\centering
\includegraphics[width=0.9\linewidth]{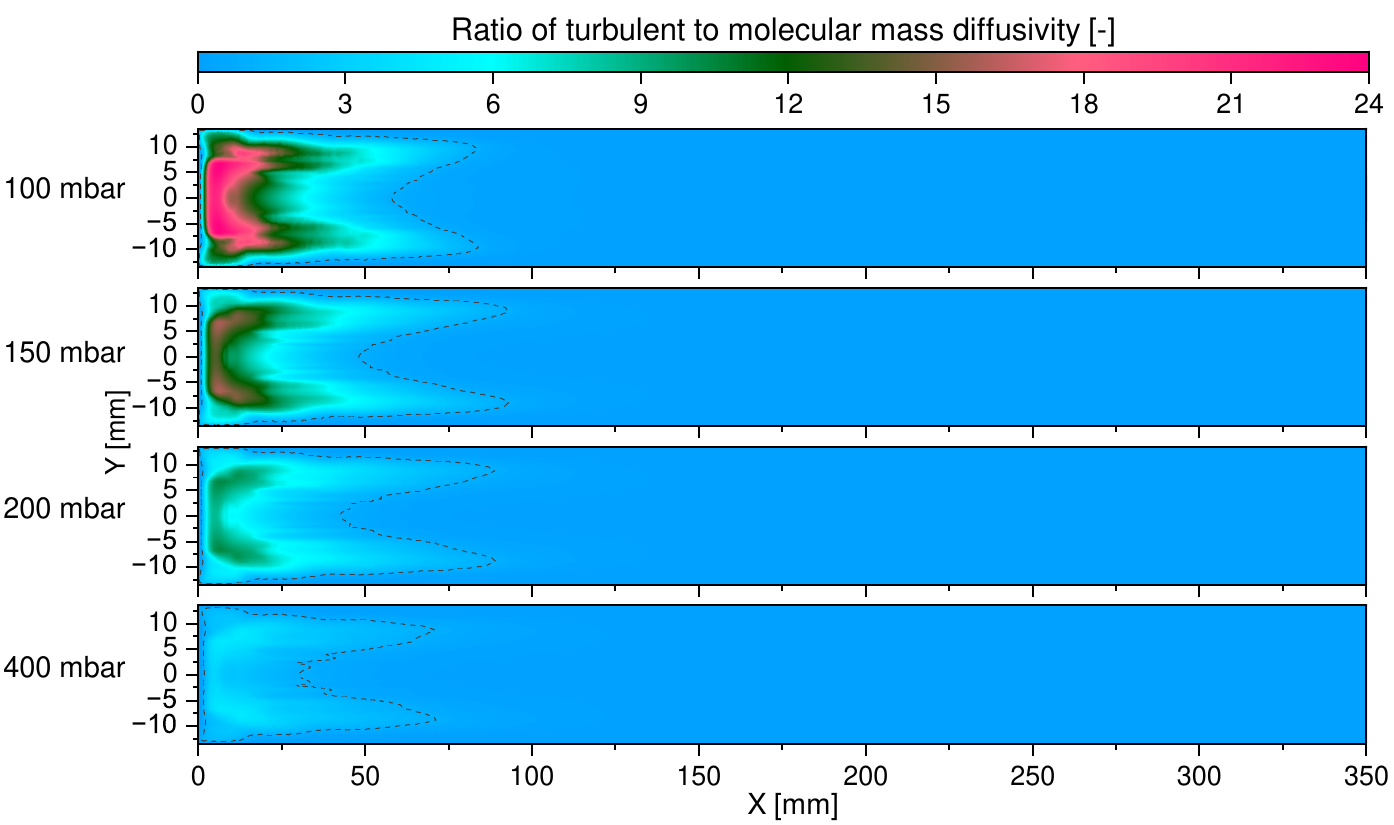}
\caption{Ratio of turbulent to molecular mass diffusivity in a cross-sectional plane at the mid-height of the reactor under different pressure conditions. \Lex{Only the axial region from 0 to 350 mm is shown.}  \RVt{The dashed line denotes the boundary of 1.}}
\label{fig:tur_diffusion_different_pressure}
\end{figure}

\begin{figure}[h]
\centering
\includegraphics[width=0.9\linewidth]{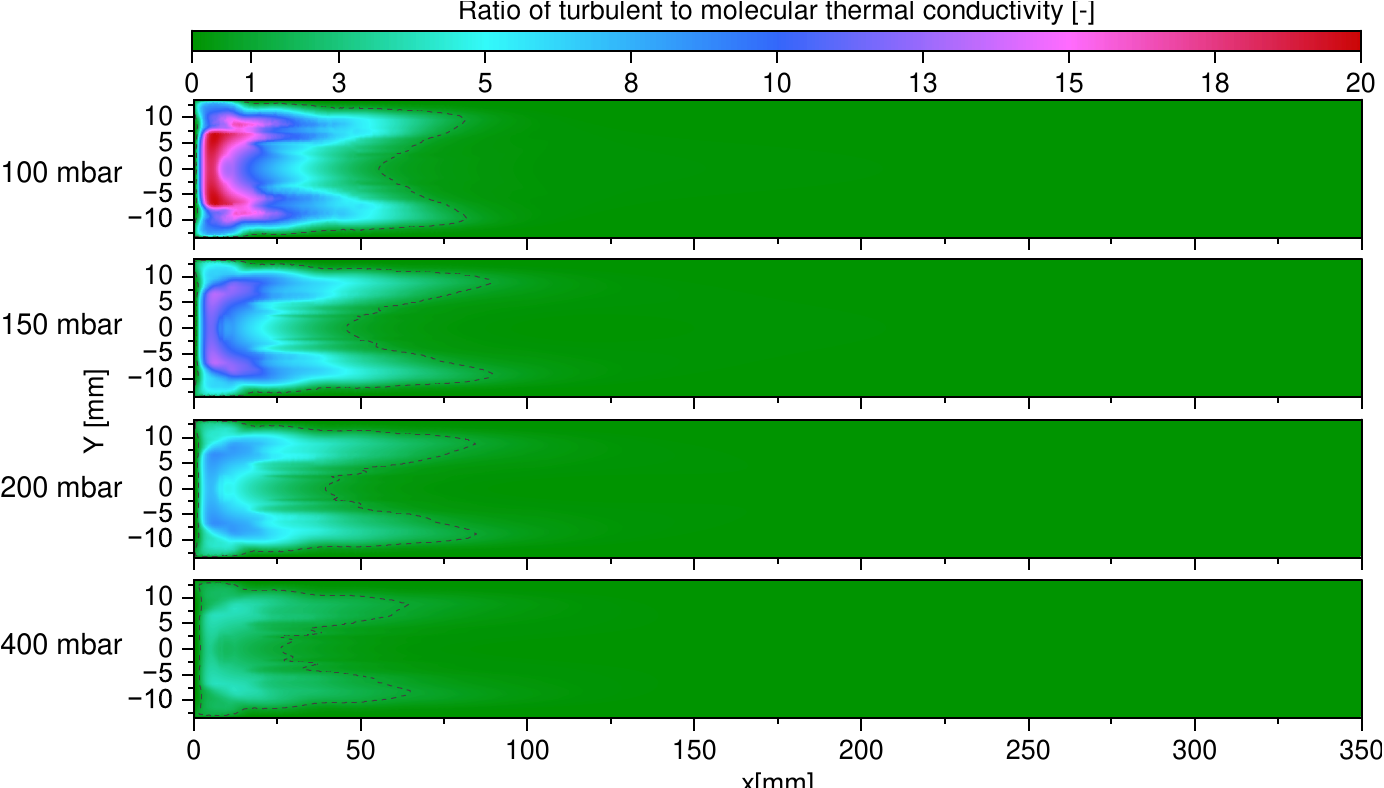}
\caption{Ratio of turbulent to molecular thermal conductivity in a cross-sectional plane at the mid-height of the reactor under different pressure conditions. \Lex{Only the axial region from 0 to 350 mm is shown.}  \RVt{The dashed line denotes the boundary of 1.}}
\label{fig:tur_cond_different_pressure}
\end{figure}

\RVt{To evaluate the effect of turbulence, Fig.~\ref{fig:tur_diffusion_different_pressure} and  Fig.~\ref{fig:tur_cond_different_pressure} compare turbulent and molecular transport using \(k_t/k_m\) and \(D_t/D_m\). The molecular transport properties used in the current CFD model are calculated from the local mixture composition using the transport-property formulations described above. However, the CFD model does not directly provide the representative mixture-averaged molecular diffusivity \(D_m\) as a standard post-processing output for this comparison. Therefore, \(D_m\) is estimated here only for diagnostic post-processing by assuming a Lewis number value \(Le=1\), a practical approximation widely used in combustion modeling \cite{mukundakumar2021new, poinsot2005theoretical}, so that \(D_m\approx k_m/(\rho c_p)\).}

\RVt{These results show that turbulence is mainly confined to the upstream reactor region near the reactor top, above the plasma zone, with only limited effects elsewhere. }  This spatial localization indicates that turbulence is not primarily driven by thermal expansion caused by plasma heating. Instead, it originates from the disturbed flow generated by the reactor geometry and inlet configuration. Recirculating gas reaches the reactor top and is redirected by wall confinement. The redirected flow then interacts with the high-velocity tangential inflow, producing strong local mixing, velocity gradients, and shear layers. \RVt{Their highest values also occur near the reactor top and decrease rapidly downstream, confirming that turbulence is generated mainly in the upstream recirculation and inlet-interaction region rather than throughout the full reactor volume. Additional turbulence quantities in the Supporting Information show that the turbulent Reynolds number, turbulent viscosity ratio, and turbulence intensity follow the same spatial trend.}

\RVt {Turbulence is strongest at 100 mbar and decreases markedly as pressure rises. This trend is mainly caused by the pressure-dependent gas velocity, as shown in Fig.~S3 in the Supporting Information. At a fixed mass flow rate, lower pressure requires a higher inlet velocity, which strengthens velocity gradients and shear layers when the incoming flow interacts with the recirculation region. At higher pressures, the larger gas density reduces the axial velocity, weakens the shear layer, and stabilizes the flow. The reduced turbulent viscosity ratio, turbulent Reynolds number, turbulence intensity, and turbulent-to-molecular transport ratios at higher pressures therefore all indicate a weaker turbulent contribution. This pressure-driven weakening of turbulence is consistent with the atmospheric-pressure simulations of Van Poyer \textit{et al.}, who also reported limited turbulence effects at a similar flow rate of 10~slm \cite{van2025influence}.}

{
\color{black}
\subsection{Cooling-controlled axial CO transport and design implications}

To connect the axial production and loss of CO with the local thermal history, we consider two complementary quantities. The first is the net axial molar flow rate of species \(i\), obtained by integrating its axial molar flux over each cross-sectional plane:
\begin{equation}
\phi_i(x)=22.414\times60\times2\pi
\int_0^{r_{tube}} n_i(x,r)u_x(x,r)r\,dr,
\label{eq:flow_rate}
\end{equation}
where \(n_i\) [mol m\(^{-3}\)] is the molar concentration and \(u_x\) [m s\(^{-1}\)] is the axial velocity. The radial coordinate is \(r=\sqrt{y^2+z^2}\), and \(r_{tube}\) denotes the reactor-tube radius. Thus, \(\phi_i\) [slm] represents the net molar flow through the entire tube cross-section. Because the radial flux distribution is integrated out, axial changes in \(\phi_i\) reflect net species production or consumption rather than transverse redistribution. \RVt{It should be noted that, in the upstream recirculation region, the axial velocity changes sign between the reactor core and the outer flow region. As a result, the CO flux carried by the recirculating core flow and that carried by the surrounding forward flow have opposite signs in Eq.~(\ref{eq:flow_rate}). A near-zero upstream value of \(\phi_{CO}\), as observed at lower pressures in Fig.~\ref{fig:net_CO_flow}, therefore does not imply the absence of CO in the upstream region. Instead, it indicates that upstream CO transport by recirculation is nearly balanced by CO transport in the opposite direction through the outer flow region. }

\begin{figure}[h]
\centering
\includegraphics[width=0.5\linewidth]{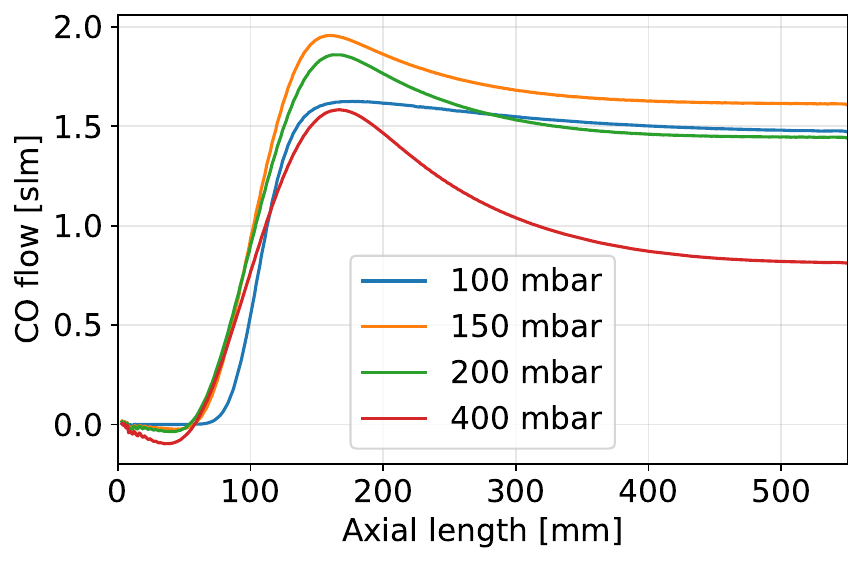}
\caption{Net axial CO molar flow rate as a function of axial position at different pressures, calculated using Eq.~(\ref{eq:flow_rate}).}
\label{fig:net_CO_flow}
\end{figure}
 
The second analysis is the axial temperature-change rate along the reactor centerline:
\begin{equation}
c_{axial}=-u_x(x)\frac{dT}{dx},
\label{eq:cooling_rate}
\end{equation}
where \(dT/dx\) [K m\(^{-1}\)] is the axial temperature gradient. Positive \(c_{axial}\) indicates cooling along the flow direction, whereas negative values indicate local reheating. Together, \(\phi_{CO}\) and \(c_{axial}\) link the observed CO transport to pressure-dependent thermal relaxation.
 
\begin{figure}[h]
\centering
\includegraphics[width=0.5\linewidth]{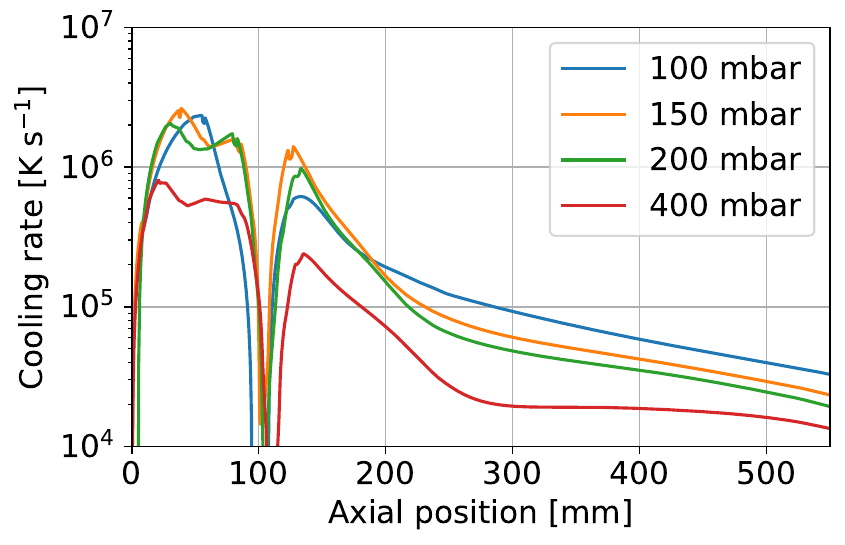}
\caption{Axial cooling rate along the reactor centerline at different operating pressures, calculated using Eq.~(\ref{eq:cooling_rate}).}
\label{fig:cooling_rate}
\end{figure}
 
For all investigated pressures, the CO flow increases rapidly within the plasma region (Fig.~\ref{fig:net_CO_flow}). This increase reflects strong CO$_2$ dissociation in the high-temperature plasma region. The flow generally reaches a maximum shortly downstream of the plasma, at an axial position of approximately 160~mm. At 100~mbar, no distinct local maximum develops, despite a plasma-core temperature of approximately 4000~K, which is close to the optimal temperature of CO$_2$ dissociation. This behavior is consistent with faster gas transport through the CO-production region. The highest peak CO flow occurs at 150~mbar, followed by only a moderate downstream decrease.
 
The stagnation position separates the upstream recirculation path from the downstream afterglow path (Fig.~\ref{fig:v_T_150mbar} and S7). Near this position, \(u_x\) approaches zero and axial convective heat removal becomes weak. Consequently, local power deposition and heat transport from the surrounding hot gas can produce negative values of \(c_{axial}\) around 100~mm (Fig.~\ref{fig:cooling_rate}). Upstream of the stagnation position, gas is redirected toward the reactor top through the recirculation region. Cooling along this path is generally faster than in the downstream region, owing to stronger turbulence near the reactor top (Fig.~\ref{fig:tur_diffusion_different_pressure} and \ref{fig:tur_cond_different_pressure}). This rapid thermal relaxation limits the upstream extent of the high-temperature region and helps preserve the recirculated CO, especially at lower pressures (Fig.~\ref{fig:cooling_rate}).
 
As pressure increases, the reduced turbulent contribution weakens upstream cooling and allows CO recombination within the upper recirculation region (Fig.~\ref{fig:tur_diffusion_different_pressure} and Fig.~\ref{fig:tur_cond_different_pressure}). At 400~mbar, the net CO flow becomes negative near \(x=50\)~mm (Fig.~\ref{fig:net_CO_flow}). This sign indicates that reverse CO transport in the reactor core exceeds forward transport in the outer region. The recirculation structure and the negative local CO-production region in Fig.~\ref{fig:net_CO_production_different_pressure} support partial consumption of the upstream-transported CO. The magnitude may depend on numerical diffusion and mesh resolution. However, second-order spatial discretization was applied to the momentum, energy, and species equations. The mesh-independence analysis also showed limited variation in the principal flow, temperature, and CO$_2$ conversion fields. The predicted upstream transport is therefore unlikely to arise primarily from numerical or grid-induced errors.
 
Downstream of the plasma, \(c_{axial}\) decreases with axial distance and shows a strong pressure dependence. At 100~mbar, rapid cooling restricts the high-temperature afterglow and limits CO recombination. Increasing pressure slows the cooling process and maintains elevated temperatures over a longer downstream distance. This effect is most pronounced at 400~mbar. Combined with more frequent three-body collisions, the extended hot afterglow promotes stronger CO recombination. As a result, more than 60\% of the produced CO is lost at 400~mbar.
 
The relative flow changes further distinguish the two loss pathways. At 400~mbar, approximately 0.15~slm of CO flow is lost in the upstream recirculation region, compared with approximately 0.8~slm in the downstream afterglow. The upstream contribution therefore represents about 16\% of the combined loss in these regions. It is measurable but remains secondary to downstream recombination. At lower pressures, the upstream CO flow remains close to zero, indicating that most recirculated CO is preserved and subsequently redirected downstream. This preservation is particularly evident at 100~mbar.

The present results provide three practical guidelines for suppressing CO recombination and improving CO$_2$ conversion. First, rapid quenching becomes increasingly important with increasing pressure. The afterglow cooling rate decreases as pressure rises, prolonging the exposure of CO to high-temperature conditions and increasing recombination losses. An external rapid-quenching stage positioned immediately downstream of the plasma can therefore improve CO preservation, particularly at elevated pressures. This conclusion is consistent with previous experimental studies, showing enhanced CO$_2$ conversion through intensified quenching under high-pressure conditions \cite{van2024effluent,mercer2023post}. Second, the reactor-top flow field should be optimized to maintain sufficient turbulent mixing and cooling in the upstream recirculation region. At low pressure, stronger turbulence near the reactor top increases the local cooling rate and helps preserve CO. At higher pressure, this turbulent contribution weakens, and a fraction of CO flow is lost in the reactor-top recirculation region. Optimizing the tangential inlet configuration or upper-reactor geometry to strengthen controlled turbulent mixing can therefore reduce upstream CO loss. Third, the operating pressure should be selected to balance plasma contraction and product preservation. In the present work, 150~mbar provides this balance, because CO$_2$ dissociation is strongly activated while downstream CO recombination remains comparatively limited.

}

\section{Conclusions}

\RVt{In this work, a three-dimensional thermochemical CFD model was developed for a vortex-stabilized microwave CO$_2$ plasma reactor operated over the pressure range of 100–400 mbar. The model used experimentally constrained plasma sizes and volumetric heat-source distribution and solved finite-rate heavy-particle chemistry for a multi-component CO$_2$/CO/O/O$_2$/C mixture in the steady state. The model was validated against measured radial gas-temperature profiles and post-plasma CO$_2$ conversion. Among the tested chemical mechanisms, the GRI-Mech-based mechanism gave the best agreement with the experimental pressure trend, including the conversion maximum at 150~mbar and the decrease at 400~mbar.}

\RVt{The non-monotonic pressure dependence of CO$_2$ conversion arises from a balance between enhanced CO$_2$ dissociation at intermediate pressure and increasing CO recombination at higher pressure. At 100~mbar, the discharge is wider, and the gas velocity is higher. Although the plasma reaches temperatures favorable for CO$_2$ dissociation, the lower heavy-particle collision frequency and faster convective transport limit the net conversion. Increasing the pressure to 150~mbar contracts the emission-based plasma region, increases the local gas temperature, and strengthens the dominant CO-forming pathways, while downstream CO loss remains comparatively limited. This gives the highest conversion in the present reactor. At 400~mbar, the plasma remains confined and axially elongated, but the higher gas density and slower afterglow cooling promote three-body CO recombination. As a result, more than 60\% of the CO formed near the plasma is lost downstream.}

\RVt{The reaction analysis shows that CO formation is mainly governed by direct CO$_2$ dissociation and O-assisted CO$_2$ conversion, but these two pathways are spatially separated. Direct CO$_2$ dissociation is strongest in the high-temperature plasma core, whereas the O-assisted pathway contributes mainly near the edges of the high-temperature plasma region and in the surrounding high-temperature region. The reactor-integrated reaction analysis further shows that these two pathways make comparable contributions at 100--150~mbar. At higher pressures, however, three-body CO recombination increasingly drives the reverse direction of the direct dissociation pathway, causing its net contribution to decrease more strongly than that of the O-assisted pathway.}

\RVt{The flow analysis shows that transport is strongly shaped by the reactor-top recirculation structure. This recirculation region redistributes part of the CO-rich gas upstream, while the interaction between the redirected flow and the incoming tangential gas stream generates localized turbulent transport near the reactor top. The transport analysis shows that turbulent heat and mass transport can locally exceed molecular transport in this upstream region. In most of the downstream reactor, however, molecular transport remains dominant. With increasing pressure, the higher gas density lowers the gas velocity, weakens shear generation, and reduces the turbulent contribution to heat and species transport.}

\RVt{These results suggest three practical optimization directions for microwave CO$_2$ plasma reactors. First, high-pressure operation should be combined with stronger downstream quenching to suppress afterglow CO recombination. Second, the reactor-top flow field should be optimized to strengthen turbulent mixing and cooling in the upstream recirculation region. This can reduce the partial CO loss observed there at higher pressure and improve retained CO$_2$ conversion. Third, pressure should be selected to balance plasma contraction and product preservation. In the present geometry, 150~mbar provides this balance.}

\RVt{Future work should focus on making the model more self-consistent and more predictive. First, coupling the electromagnetic field to the thermochemical flow model would allow the power-deposition profile and plasma size to be calculated rather than prescribed from experiments. Second, Lagrangian particle tracking or an age-of-fluid analysis should be used to quantify residence-time distributions in the three-dimensional recirculating flow. Third, extended chemical mechanisms including additional species and surface recombination could be tested to assess their influence on the downstream species balance. These developments would further improve the applicability of the model for reactor optimization and scale-up.}


\section{CRediT authorship contribution statement}
\textbf{Qinghao Shen}: Writing –- Original draft, Conceptualization, Validation, Formal analysis, Investigation, Methodology, Software, Visualization. \textbf{Cas van Deursen}: Data curation, Visualization, Writing –- review and editing. \textbf{Pieter Willem Groen}: Software, Methodology, Writing –- review and editing. \textbf{Lex Kuijpers}: Visualization, Writing –- review and editing. \textbf{Mauritius C.M. van de Sanden}: Funding acquisition, Resources, Supervision, Project administration, Writing – review and editing.

\section{Declaration of competing interest}
The authors declare that they have no competing interests.

\section{Data availability}
Data will be made available on request.

\newpage

\clearpage
\section*{Supporting Information}

\setcounter{figure}{0}
\setcounter{table}{0}
\setcounter{equation}{0}
\renewcommand{\thefigure}{S\arabic{figure}}
\renewcommand{\thetable}{S\arabic{table}}
\renewcommand{\theequation}{S\arabic{equation}}
\renewcommand{\theHequation}{S\arabic{equation}}

  The dynamic viscosity (\(\mu\) [Pa s]) is defined based on kinetic theory as: 
\begin{equation}
    \mu= \sum_i \frac{X_i \mu_i}{\sum_jX_j\phi_{ij} }
\end{equation}
where \( X_i\) [dimensionless] is the specific species molar fraction, the function \(\phi_{ij}\) is calculated by:
\begin{equation}
\phi_{ij}=\frac{ \left[ 1+  \left(\frac{\mu_i}{\mu_j}\right)^{0.5} \left(\frac{M_{j} }{M_{i}}\right)^{0.25} \right]^{2} }{\left[     8\left(1 + \frac{ M_{i}}{M_{j}}\right)\right]^{0.5}}
\end{equation}
where \(\mu_i\) [Pa s] is the specified species viscosity, which is defined as:
\begin{equation}
\mu_i= 2.67\times 10^{-6} \frac{\sqrt{M_iT}}{\sigma_i^2 \RVt{\Omega_{i}^{(2,2)*}}}
\end{equation}
where  \(M_i\) [kg mol$^{-1}$] is the specific molecular weight, \(T\) [K] is the gas temperature,  \(\sigma_i\) and \RVt{\(\Omega_{i}^{(2,2)*}\)} are the corresponding Lennard-Jones parameter and the \RVt{viscosity} diffusion collision integral, \Lex{respectively}.

\(\overline{\mathbf{u}'\otimes\mathbf{u}'}\) is the Reynolds stress term, whose number depends on the particular choice of the turbulence model \cite{antonini2018analysis}.  The Boussinesq hypothesis is commonly used to connect the Reynolds stresses with the mean velocity gradients, offering the advantage of a relatively low computational cost for evaluating the turbulent viscosity \cite{matsson2022introduction}: 
\begin{equation}
   -\rho \overline{\mathbf{u}'\otimes\mathbf{u}'}=\mu_t\left( \nabla \mathbf{u} + (\nabla \mathbf{u})^ \mathrm{T} \right)     -  \tfrac{2}{3}( \rho k + \mu_t \nabla\cdot\mathbf{u})\mathbf{I}
\end{equation}
where \(k\) [m$^2$s$^{-2}$] is the turbulent kinetic energy,  \(\mu_t\) [Pa s] is the turbulent viscosity.

The turbulence kinetic energy \(k\) and the specific dissipation rate \(\omega\) [s$^{-1}$] are calculated by:

\begin{equation}
   \nabla \cdot (\rho\, k \,\mathbf{u}) 
= \nabla \cdot (\Gamma_k \nabla k) + G_k - Y_k + S_k + G_k
\end{equation}
\begin{equation}
   \nabla \cdot (\rho \, \omega \, \mathbf{u}) 
= \nabla \cdot (\Gamma_\omega \nabla \omega) + G_\omega - Y_\omega + S_\omega + G_\omega
\end{equation}
where \(G_k\) [kg m$^{-1}$s$^{-3}$] and \(G_{\omega}\) [kg m$^{-3}$s$^{-2}$] represent the generation of \(k\) and \(\omega\) due to mean velocity gradients. \(\Gamma_k\) and \(\Gamma_{\omega}\)   \Lex{(both in [m$^2$s$^{-1}$])} denote the effective diffusivity of \(k\) and \(\omega\), respectively. \(Y_k\) [kg m$^{-1}$s$^{-3}$] and \(Y_{\omega}\) [kg m$^{-3}$s$^{-2}$]  are the dissipation of \(k\) and \(\omega\) due to turbulence. \(S_k\) [kg m$^{-1}$s$^{-3}$] and \(S_{\omega}\) [kg m$^{-3}$s$^{-2}$] are the source terms. \(G_k\) [kg m$^{-1}$s$^{-3}$] and  \(G_\omega\) [kg m$^{-3}$s$^{-2}$] account for buoyancy terms. A more detailed explanation of all the above terms can be found in the \Lex{ANSYS} Fluent guidance booklet \cite{matsson2022introduction}. \RVt{Compared with using the standard $k$--$\epsilon$ or $k$--$\omega$ model alone, the SST model generally provides a more reliable eddy-viscosity prediction for near-wall flows and flows with strong shear or separation, because it blends the near-wall behavior of the $k$--$\omega$ model with the free-stream robustness of the $k$--$\epsilon$ model and includes a shear-stress limiter in the eddy-viscosity formulation \cite{menter1994two}.} The turbulent viscosity in the SST model is derived by:

\begin{equation}
\mu_t = \frac{\rho \, k}{\omega} \; \frac{1}{\max\Big( \frac{1}{\alpha^*}, \frac{S F_2}{\alpha_1 \, \omega} \Big)}
\end{equation}
where \(S\) [s$^{-1}$] is the strain rate magnitude,  \(\alpha^*\) [dimensionless] is defined as a damp factor for turbulent viscosity \cite{matsson2022introduction}. \(\alpha_1\) [dimensionless] is constant number, fixed as 0.31, and \(F_2\) [dimensionless] is transition function, which equals to tanh(\(\Phi_2^2\)). Here, \(\Phi_2\) is calculated by:
\begin{equation}
\Phi_2 = \max \Bigg[ \frac{2 \sqrt{k}}{0.09 \, \omega \, y}, \;\; \frac{500 \, \mu}{\rho \, y^2 \, \omega} \Bigg]
\end{equation}
where \(y\) [m] denotes the distance to the nearest next surface.

\subsection{Transport of species}
 \(D_{ij}\) [m$^2$s$^{-1}$] is the binary mass diffusion coefficient of species \textit{i} into species \textit{j}, which is computed by modification of the Chapman-Enskog formula:
\begin{equation}
D_{ij} =
\frac{3}{16 \, p \, \sigma_{ij}^{2}} 
\; \sqrt{ \frac{2 \pi k_B^3 T^3}{m_{ij}} } \;
\frac{1}{ \RVt{\Omega^{(1,1)*}}},
\label{eq:Dij_SI}
\end{equation}
where \(m_{ij}\) [dimensionless] is the reduced mass, \(\sigma_{ij}\) [m] is the reduced collision diameter, which is equal to the average value of two Lennard-Jones collision diameters \(\sigma_{i}\) and \(\sigma_{j}\) (both in [m]). \RVt{\(\Omega^{(1,1)*}\)}~[dimensionless] is the non-dimensional diffusion collision integral, which is a measure of the interaction of the molecules in the system. It is  approximated in Ansys Fluent as \cite{matsson2022introduction}:
\begin{equation}
  \RVt{ \Omega^{(1,1)*}} = (T^*_D)^{-0.145} +(T_D^* +0.5) ^{-2}
\end{equation}
where \(T^*_D\) [dimensionless] is the dimensionless temperature, computed by: 
\begin{equation}
T_{D}^{*} = \frac{k_{B} T}{\varepsilon_{ij}}
\end{equation}
with \(k_B\) the Boltzmann constant, and \(\varepsilon_{ij}\) [J] denotes the Lennard–Jones potential well depth between two species.

\(D_{T_i}\) \RVt{[kg m$^{-1}$s$^{-1}$]} is the thermal (Soret) diffusion coefficient, which is calculated based on an empirically-based composition-dependent expression derived from \cite{kirkpatrick2024principles}:
\begin{equation}
D_{T_i}
= -2.59 \times 10^{-7}\, T^{0.659}
\left[
\frac{ M_{i}^{0.511} X_i }
     { \sum_{k=1}^{N} M_{k}^{0.511} X_k }
- Y_i
\right]
\left[
\frac{ \sum_{k=1}^{N} M_{k}^{0.511} X_k }
     { \sum_{k=1}^{N} M_{k}^{0.489} X_k }
\right]
\label{eq:DTi}
\end{equation}

\subsection{Heat balance equation }
\RVt{The mixture specific enthalpy $h$ [J kg$^{-1}$] is defined as:}
\begin{equation}
h=\sum_i Y_i h_i ,
\end{equation}
\RVt{where $Y_i$ is the mass fraction of species $i$ and $h_i$ is the specific enthalpy of species $i$. The species enthalpy is evaluated from the standard formation enthalpy and the sensible enthalpy,}

 Viscous stress tensor (\(\bar{\tau}\) [Pa]) is defined as:
\begin{equation}
\bar{\tau} = 
\mu \Bigl( 
\nabla \mathbf{u} + (\nabla \mathbf{u})^T 
- \frac{2}{3} (\nabla \cdot \mathbf{u}) \mathbf{I} 
\Bigr)
\label{eq:stress_tensor}
\end{equation}

 The thermal conductivity of each species \(k_i\)is calculated by:
\begin{equation}
  k_{i}= \frac{15}{4} \frac{\mathrm{R}}{M_i}\mu_i \bigg[ \frac{4}{15} \frac{c_{p,i}M_i}{\mathrm{R}}  + \frac{1}{3}\bigg]
\end{equation}
where R is the universal gas constant, \(c_{p,i}\) [J kg$^{-1}$K$^{-1}$] is the specific species heat capacity.

\(Q_{(x, y, z)}\) [W m$^{-3}$]  is the local volumetric heat source,  \Lex{which is determined by the experimental data in Tab.~3 in the main text. It is computed by:}
\begin{equation}
 Q_{(x, y, z)} = Q_{peak}
\exp\Bigg[-\frac{(x-(H_{plasma}/2)^2}{2\sigma_x^2} 
          -\frac{y^2}{2\sigma_y^2} 
          -\frac{z^2}{2\sigma_z^2} \Bigg]
\label{eq:gaussian_3D}
\end{equation}
where \(Q_{peak}\) [W m$^{-3}$] represents the peak volumetric power density, and is set such that the volume integral of \(Q(x, y, z)\) over the computational domain equals the prescribed total input power \cite{kotov2023validation, shen2026boosting}, the spread parameters \(\sigma_x\), \(\sigma_y\), and \(\sigma_z\) (\Lex{all} in [m]) control how concentrated or diffuse the heat source is in each direction. \Lex{Their values are derived from the experimentally measured plasma radius and length listed in Tab.~3. Specifically, the plasma radius and length are defined as the positions where the radiation intensity decreases to 50\% of its maximum value. Therefore, \(\sigma_x\), \(\sigma_y\), and \(\sigma_z\) are calculated by using the following relations:}
{
\begin{equation}
\exp\Bigg(
          -\frac{H_{plasma}^2}{8\sigma_x^2} \Bigg)= \exp\Bigg(-\frac{R_{plasma}^2}{2\sigma_y^2} 
           \Bigg)= \exp\Bigg(-\frac{R_{plasma}^2}{2\sigma_z^2} 
           \Bigg) = \frac{1}{2}
\end{equation}

}

\begin{figure}[ht]
\centering
\includegraphics[width=1\linewidth]{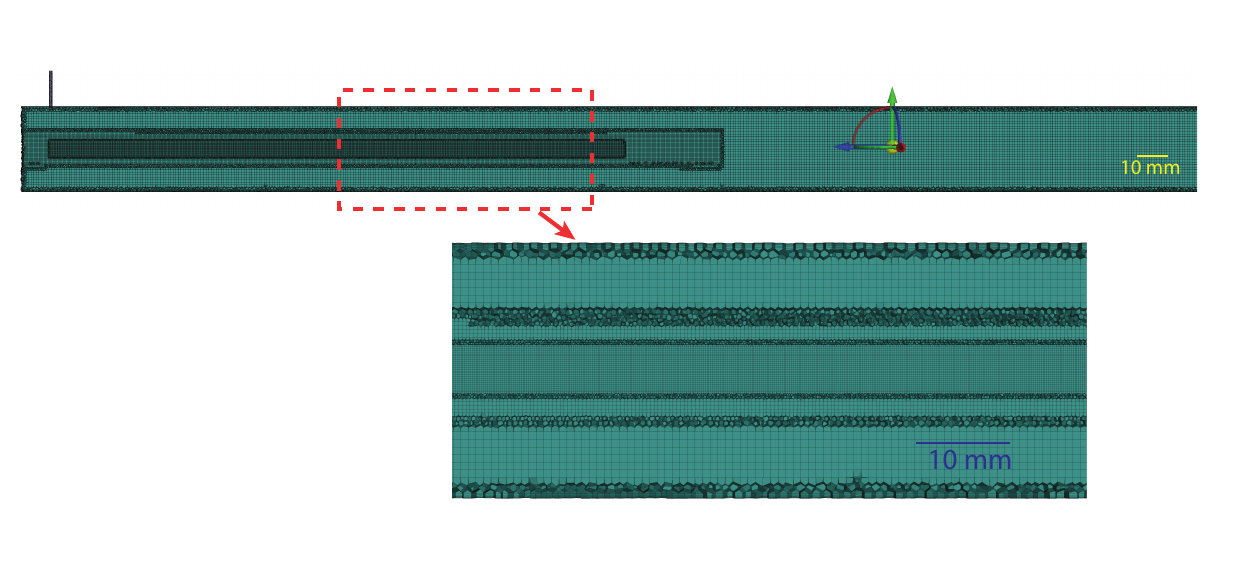}
\caption{Computational mesh over the axial range of 0–350 mm, with an enlarged view of the region indicated by the dashed box. }
\label{fig:axial_velocity}
\end{figure}

\begin{figure}[ht]
\centering
\includegraphics[width=0.5\linewidth]{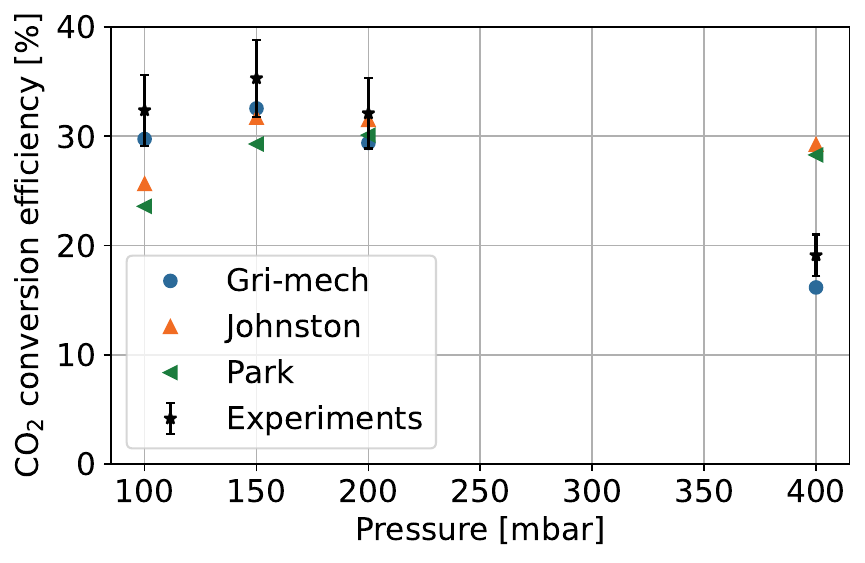}
\caption{Comparison of experimentally measured CO$_2$ conversion efficiency with model predictions obtained using different chemical mechanisms at different pressures. }

\end{figure}

\begin{figure}[ht]
\centering
\includegraphics[width=0.9\linewidth]{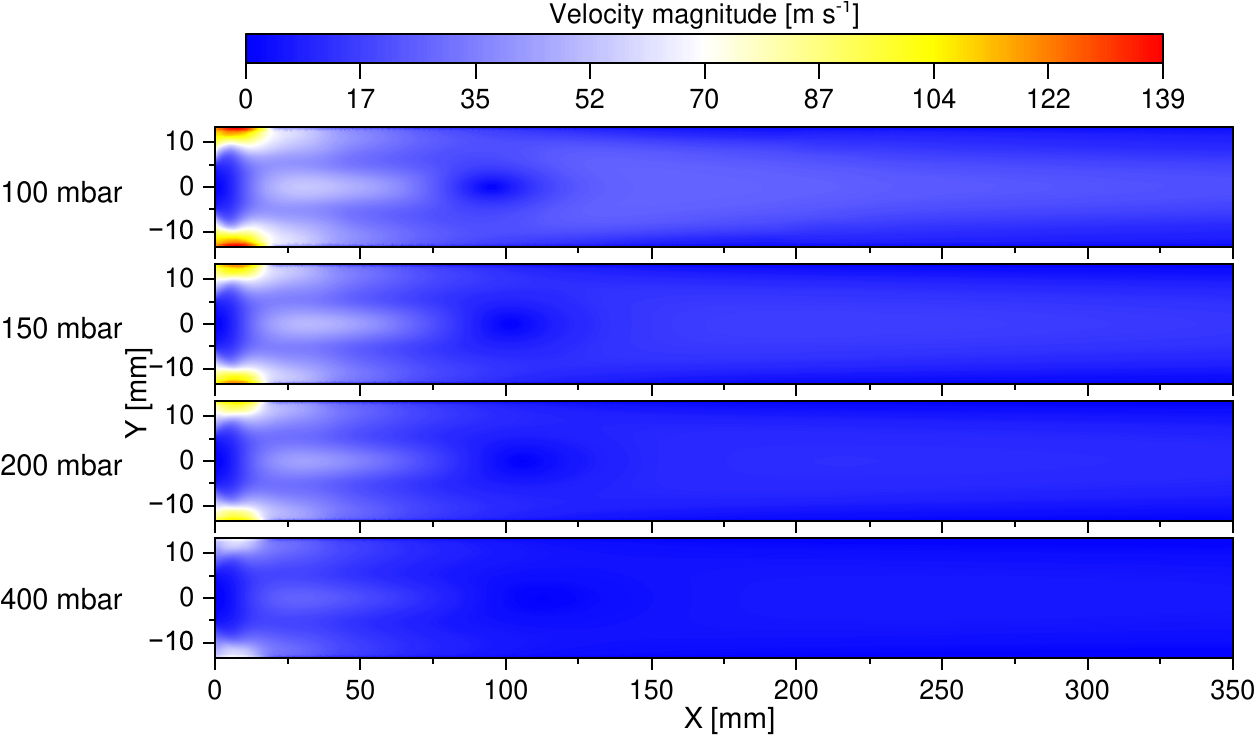}
\caption{Velocity magnitude distributions in a cross-sectional plane at the mid-height of the reactor under different pressure conditions. Only the axial region from 0 to 350 mm is shown.}

\end{figure}

\begin{figure}[ht]
\centering
\includegraphics[width=1\linewidth]{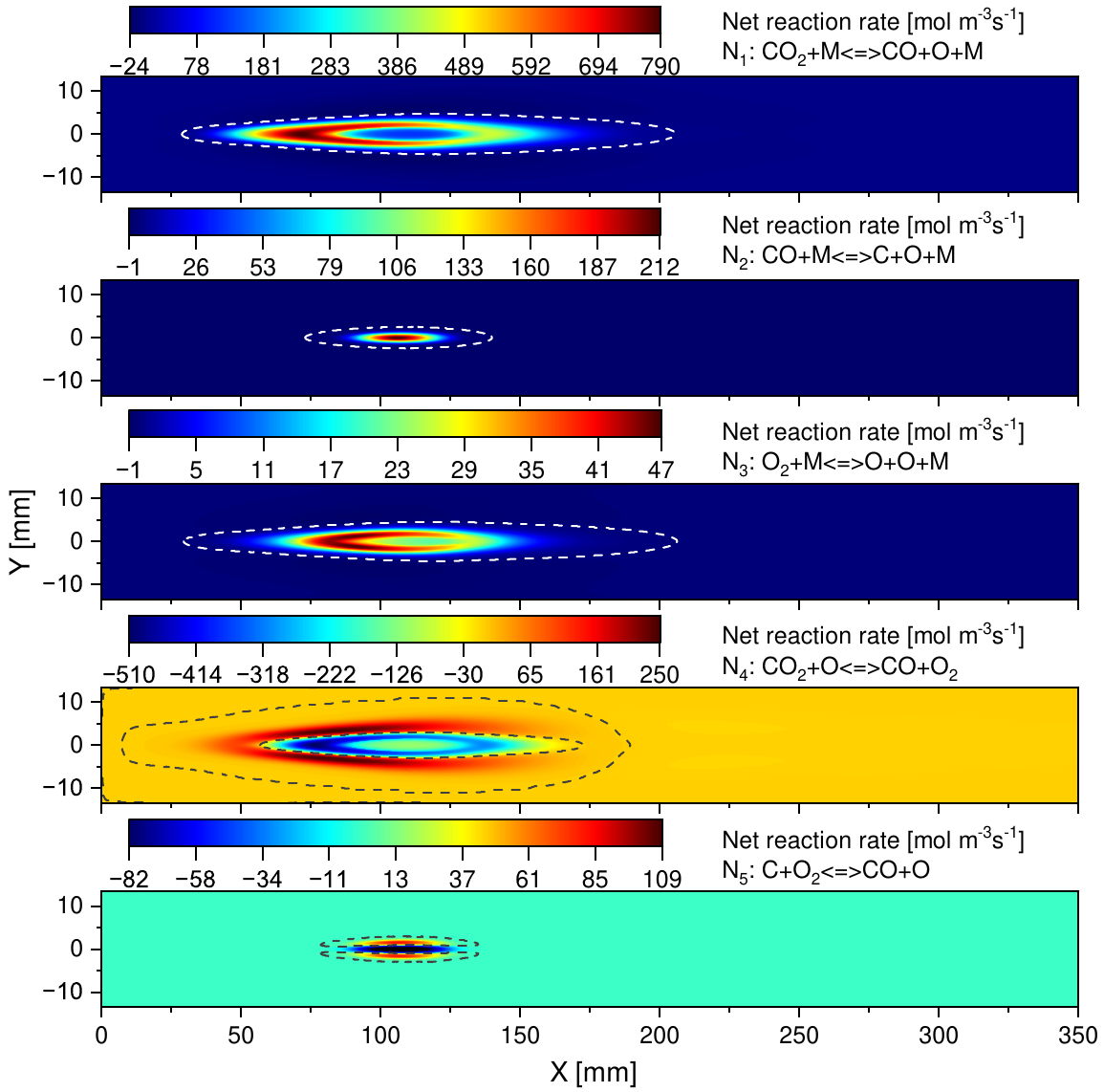}
\caption{Net reaction rate distributions in a cross-sectional plane at the mid-height of the reactor at 200 mbar. Only the axial region from 0 to 350 mm is shown. The dashed line denotes the boundary separating positive and negative net reaction rates. }
\end{figure}

\begin{figure}[ht]
\centering
\includegraphics[width=1\linewidth]{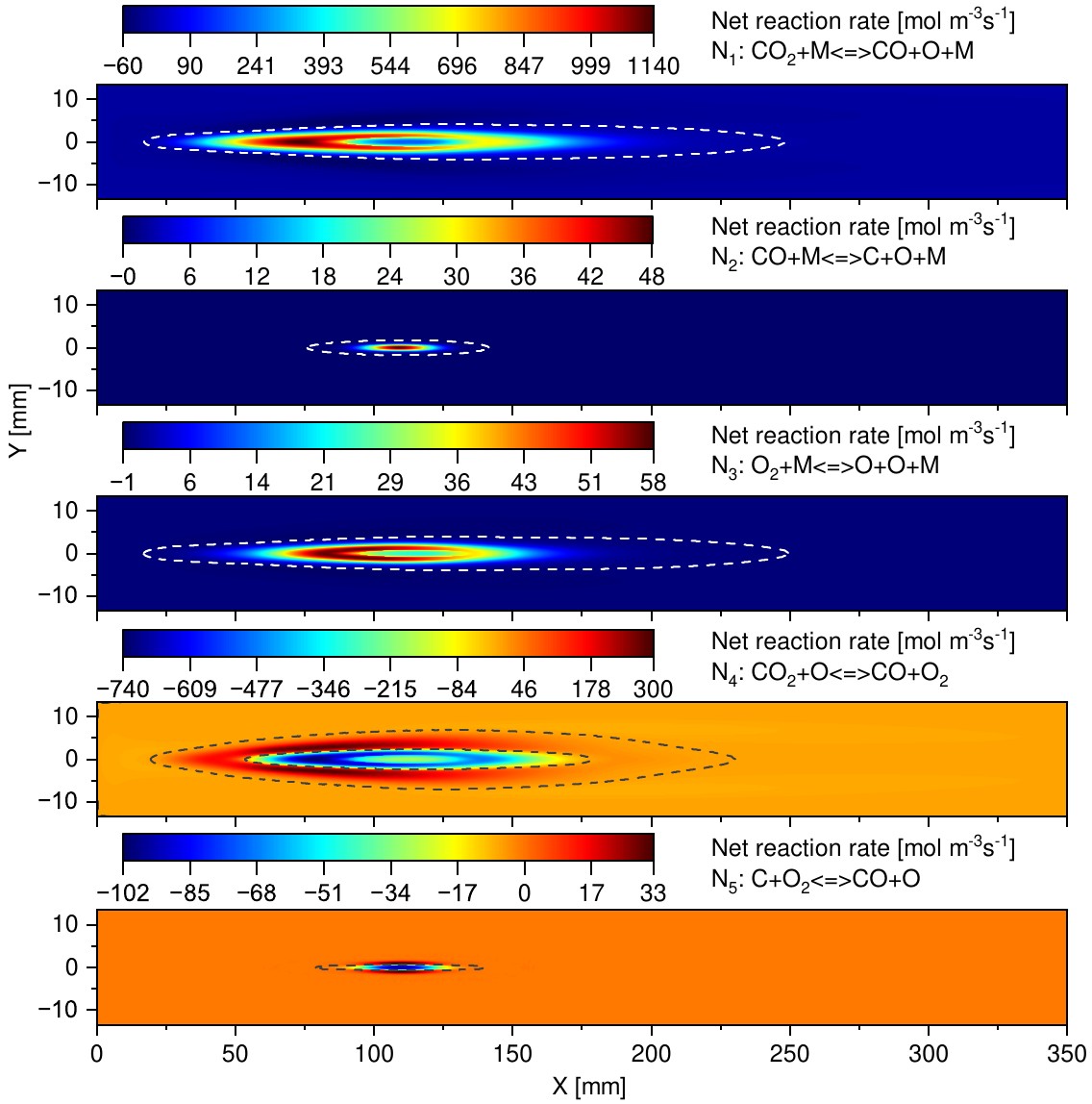}
\caption{Net reaction rate distributions in a cross-sectional plane at the mid-height of the reactor at 400 mbar. Only the axial region from 0 to 350 mm is shown. The dashed line denotes the boundary separating positive and negative net reaction rates. }
\label{fig:axial_velocity}
\end{figure}

\begin{figure}[ht]
\centering
\includegraphics[width=1\linewidth]{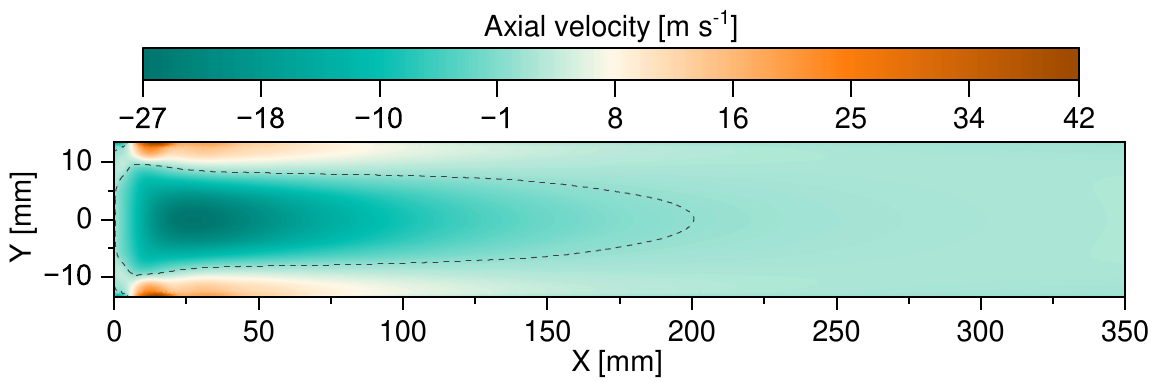}
\caption{Axial velocity distribution in the reactor mid-plane at 150 mbar, obtained using the pure-fluid model. The dashed lines indicate the locations where the axial velocity equals zero ($u_x$= 0~m~s$^{-1}$). Only the axial region from 0 to 350 mm is shown. }
\end{figure}

\begin{figure}[ht]
\centering
\includegraphics[width=1\linewidth]{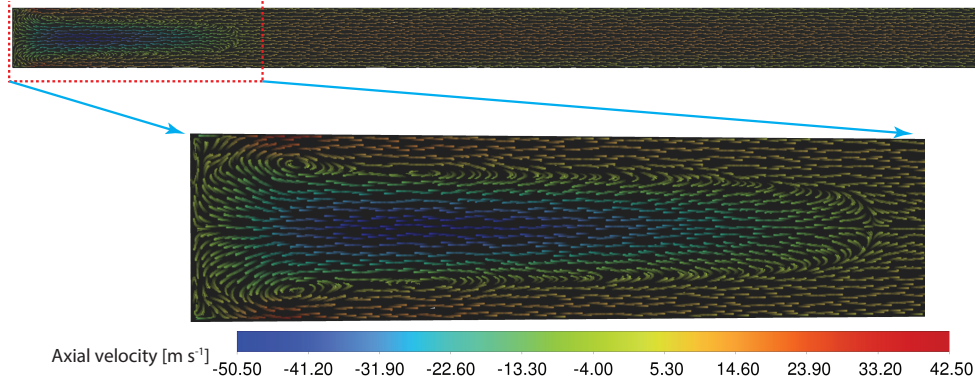}
\caption{Gas velocity distribution visualized using the Oriented Line Integral Convolution method. The orientation of the lines indicates the local flow direction, while the color represents the magnitude of the axial velocity. }
\label{fig:axial_velocity}
\end{figure}

\begin{figure}[ht]
\centering
\includegraphics[width=0.5\linewidth]{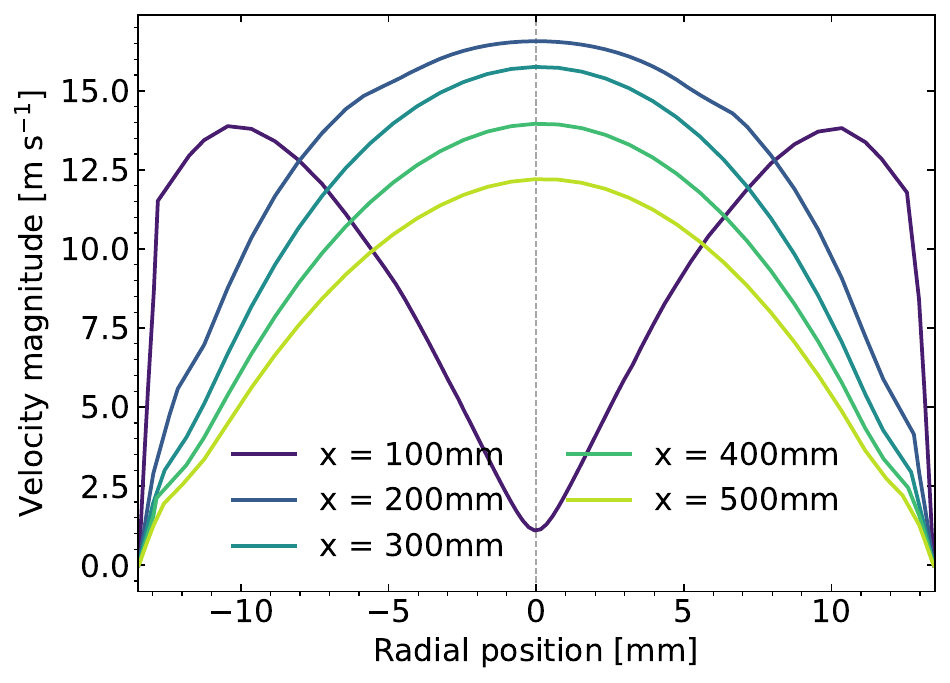}
\caption{Radial velocity magnitude at various axial positions under a pressure of 150 mbar. }
\label{fig:axial_velocity}
\end{figure}

\begin{figure}[ht]
\centering
\includegraphics[width=0.9\linewidth]{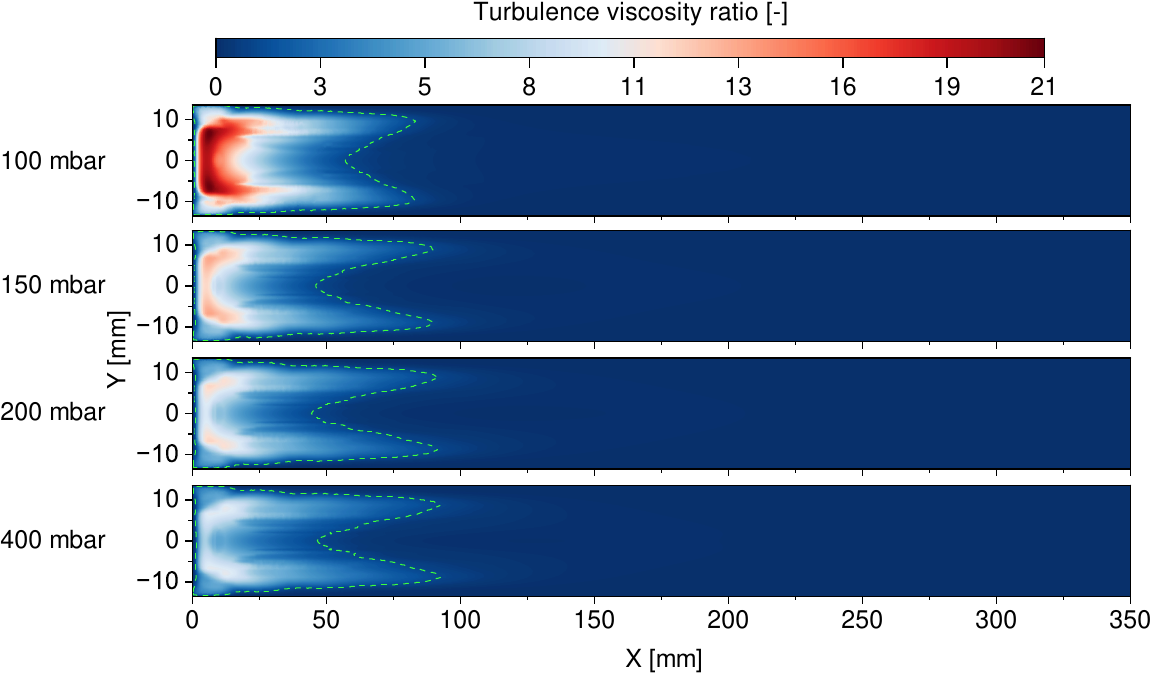}
\caption{Turbulence viscosity ratio, defined as the ratio of turbulent to molecular viscosity, distributions in a cross-sectional plane at the mid-height of the reactor under different pressure conditions. Only the axial region from 0 to 350 mm is shown.}
\end{figure}

\begin{figure}[ht]
\centering
\includegraphics[width=0.9\linewidth]{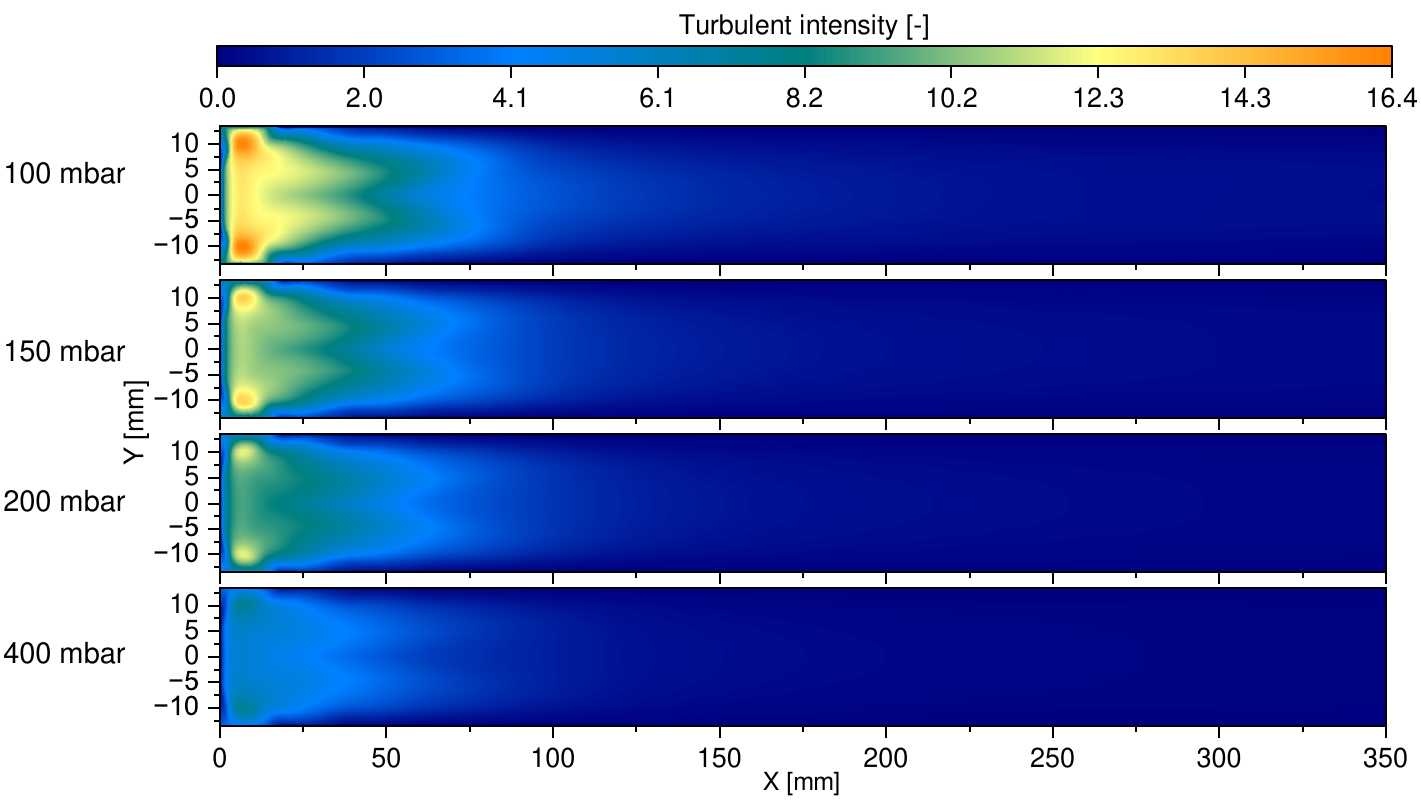}
\caption{Turbulent intensity distributions in a cross-sectional plane at the mid-height of the reactor under different pressure conditions. Only the axial region from 0 to 350 mm is shown. }
\end{figure}

\begin{figure}[ht]
\centering
\includegraphics[width=0.9\linewidth]{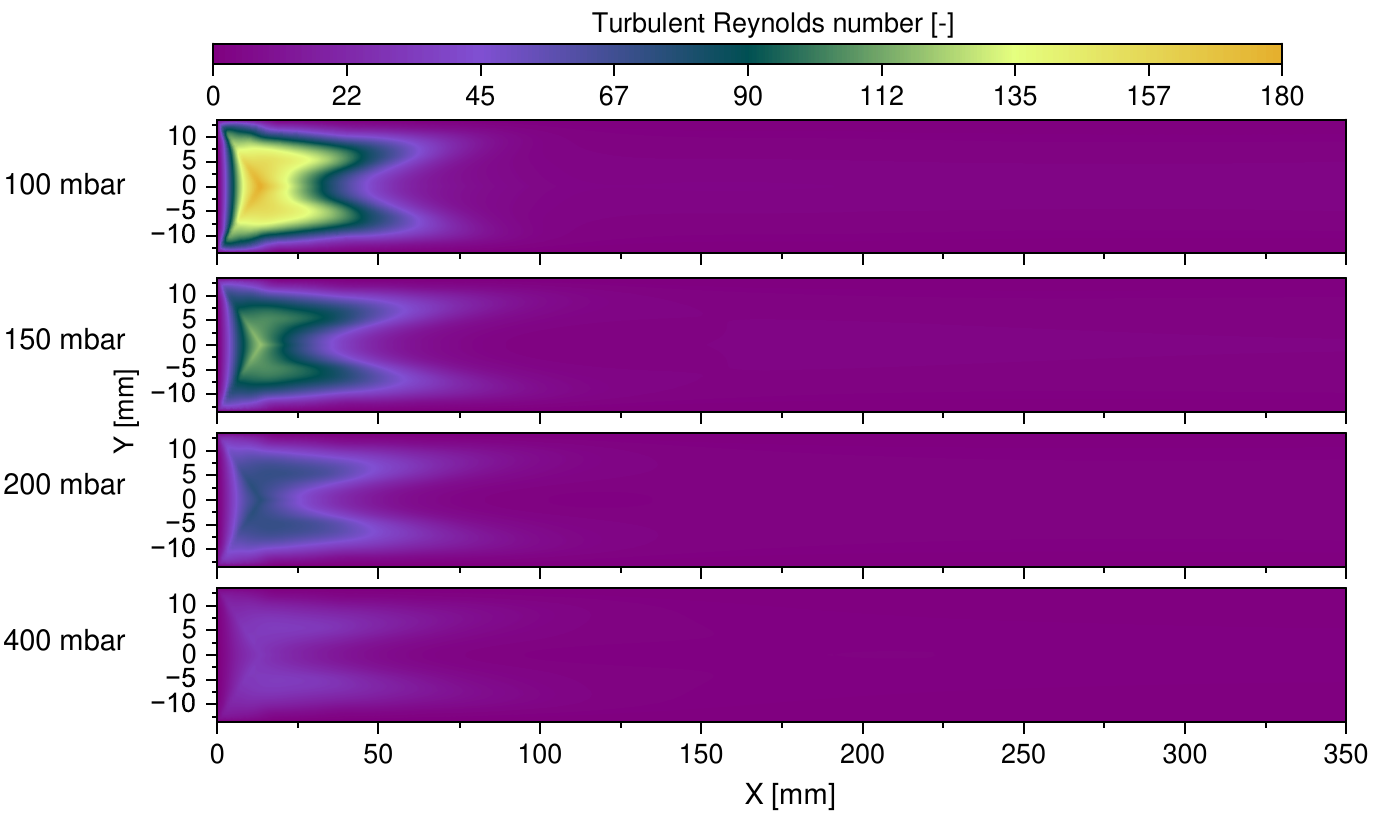}
\caption{Turbulent Reynolds number distributions in a cross-sectional plane at the mid-height of the reactor under different pressure conditions. Only the axial region from 0 to 350 mm is shown. }
\end{figure}

\clearpage

\bibliography{acs-achemso}

\end{document}